\documentclass[twocolumn,epsfig,graphics,showpacs,floatfix,noshowpacs]{revtex4}

\usepackage{amsmath,amsfonts,amssymb,graphics,graphicx,epsfig,color,times,bbm}
\usepackage{amsthm}
\usepackage{psfrag}
\usepackage{braket}
\usepackage{array}
\usepackage{multirow}
\AtBeginDocument{\usepackage{booktabs}}
\usepackage{marvosym}
\usepackage[hidelinks]{hyperref}
\hypersetup{colorlinks=false}

\begin{document}

\title{A manufacturable platform for photonic quantum computing}

\author{PsiQuantum Team}

\begin{abstract}
Whilst holding great promise for low noise, ease of operation and networking, useful photonic quantum computing has been precluded by the need for beyond-state-of-the-art components, manufactured by the millions. Here we introduce a manufacturable platform for quantum computing with photons. We benchmark a set of monolithically-integrated silicon photonics-based modules to generate, manipulate, network, and detect photonic qubits, demonstrating dual-rail photonic qubits with $99.98\% \pm 0.01\%$ state preparation and measurement fidelity, Hong-Ou-Mandel quantum interference between independent photon sources with $99.50\%\pm0.25\%$ visibility, two-qubit fusion with $99.22\%\pm0.12\%$ fidelity, and a chip-to-chip qubit interconnect with $99.72\%\pm0.04\%$ fidelity, not accounting for loss. In addition, we preview a selection of next generation technologies, demonstrating low-loss silicon nitride waveguides and components, fabrication-tolerant photon sources, high-efficiency photon-number-resolving detectors, low-loss chip-to-fiber coupling, and barium titanate electro-optic phase shifters. 
\end{abstract}

\maketitle

\section{Introduction}
\noindent It has long been understood that useful quantum computers will require error correction for fault-tolerant operation, and therefore on the order of millions of physical qubits \cite{Hollenberg2004}.
Due to their intrinsic low-noise properties, photons have been used to implement many of the foundational demonstrations of superposition, entanglement, logic gates, algorithms etc. \cite{OBrien2007}. 
However, large-scale photonic quantum computing has so far been precluded by a number of outstanding and challenging requirements. 

Since the earliest proposals for fault-tolerant optical quantum computers \cite{KLM,Ralph2001,Yoran2003,Nielsen2004,Browne2004}, it has been clear that a very large number of photonic components would be required for any useful system \cite{Ying2015,Rudolph2017}. 
Furthermore, to satisfy the requirements of error-correcting codes, these components should also perform beyond the state of the art of conventional integrated photonics \cite{Ying2015,Rudolph2017}, and must also extend outside the scope of a typical photonics library, introducing non-standard devices --- most notably high-efficiency single-photon detectors \cite{Goltsman2001,Nam2013}. 
The need for a very large number of near-identical devices motivates an emphasis on fabrication using conventional, high-volume semiconductor manufacturing processes \cite{geng2018semiconductor}. 
Finally, these devices must be integrated together into an extensive system --- demanding fast control electronics, high-power cryogenic cooling to support the operation of superconducting detectors, and low-loss, high-fidelity networking of qubits between modules. 

In this paper, we describe a technology stack and basic building blocks for photonic quantum computing, including single photon sources, waveguide-integrated superconducting single-photon detectors, single-qubit state preparation and measurement (SPAM), chip-to-chip qubit interconnects, two-photon quantum interference and two-qubit fusion; all at telecom (C band) wavelengths. These constitute the basic operations required for most approaches to photonic quantum computing \cite{KLM,Ralph2001,Yoran2003,Nielsen2004,Browne2004,Ying2015}, including fusion-based quantum computing (FBQC, recently introduced in \cite{FBQC2023}).
These components are fabricated in a commercial semiconductor foundry \cite{Giewont2019}, using a fully-integrated 300mm silicon photonics process flow, with all operations on-chip. 

In addition, we describe next-generation components including a novel spontaneous single photon source, integrated, high-efficiency photon-number-resolving detectors, low-loss silicon nitride (SiN) waveguides and components, low-loss fiber-to-chip edge coupling, and waveguide-integrated barium titanate (BTO) electro-optic phase shifters.

\begin{figure*}[ht]
\centering
\includegraphics[width=\textwidth]{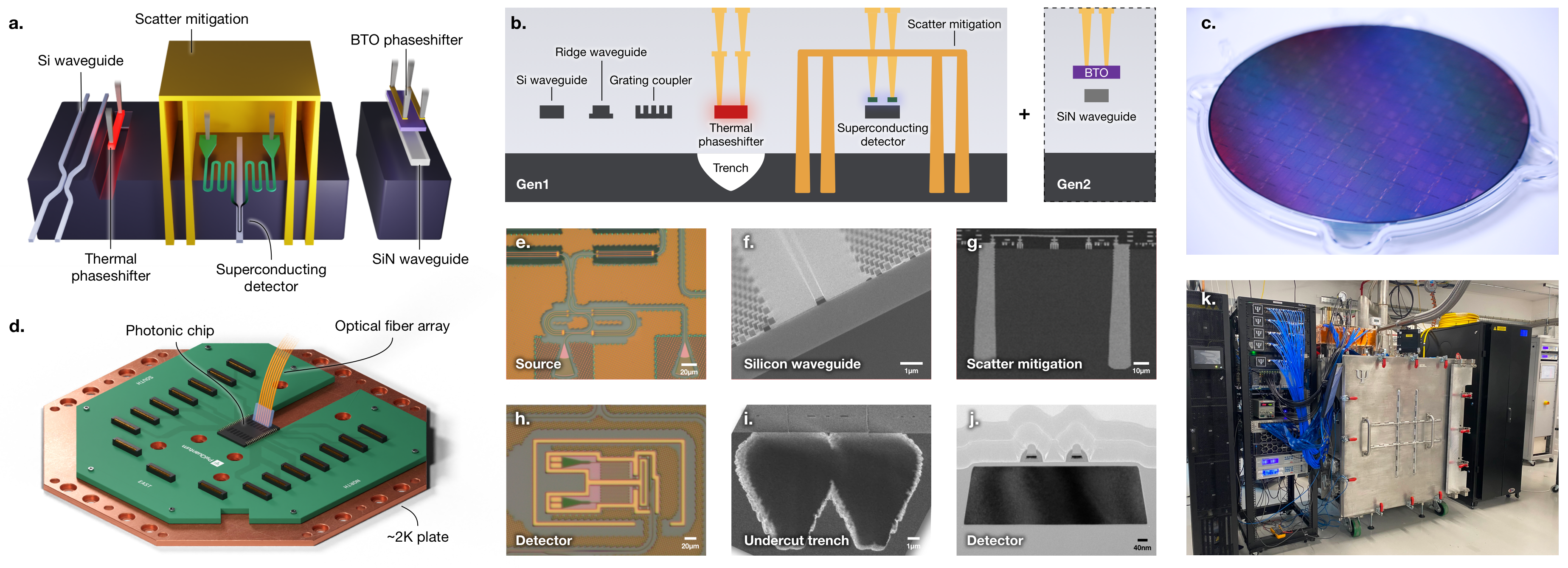}
\caption{\textbf{Manufacturable integrated quantum photonic stack.} 
\textbf{a} \& \textbf{b}, Schematic of key components and process modules. We highlight (on right) additional process steps included in our next-generation platform. 
\textbf{c}, A 300mm wafer containing single photon sources, superconducting single-photon detectors, and quantum benchmarking circuits.
\textbf{d}, A cryogenic assembly containing a photonic die, heat-spreader, electronic PCB \& 100-channel telecom fibre attach unit. 
\textbf{e-k}, optical micrograph, scanning-electron-microscope or transmission electron microscopy images of: 
\textbf{e}, photon source (top down); 
\textbf{f}, optical waveguide (cross-section); 
\textbf{g}, deep/shallow trench scattered light shield (cross-section); 
\textbf{h}, single-photon detector (top down); 
\textbf{i}, thermal-isolation trench (cross-section); 
\textbf{j}, single-photon detector on waveguide (cross-section). 
\textbf{k}, custom cryostat used in benchmarking experiments with $>\!10$ W cooling power at $2.2$ K.}
\label{fig:1}
\end{figure*}

\section{Technology stack \& building blocks}
Silicon photonics is a mature manufacturing technology, built on decades of industrial development for established applications in the communications, medical and automotive sectors \cite{Siew2020, Shekhar2024}. 
We modified an established silicon photonics manufacturing flow to include high-performance single-photon detection and photon pair generation (Fig.~\ref{fig:1}).
To our knowledge, this is the first realization of a integrated photonic technology platform capable of on-chip generation, manipulation, and detection of photonic qubits.

Our baseline quantum-photonic technology stack was developed in partnership with GlobalFoundries, and is fabricated in their 300mm state-of-the-art high-volume semiconductor foundry.
By leveraging industrial unit process steps from semiconductor manufacturing in combination with foundry design services, such as optical proximity correction and optimized process design rules, the technology inherits the scalability and performance of a high-volume commercial environment. 
The manufacturing flow includes over 20 photolithography levels and hundreds of processing and in-line measurement steps. 
Critical process modules developed include passive silicon-on-insulator (SOI) photonic waveguides, a niobium nitride (NbN) superconducting layer for single photon detection, deep metal-filled trenches for optical noise reduction, resistive heaters for phase control and optical circuit reconfigurability, grating couplers for optical input/output (I/O), back-end-of-line copper electrical interconnects, and aluminium redistribution layers. 

\begin{figure}[t]
\includegraphics[width=\columnwidth]{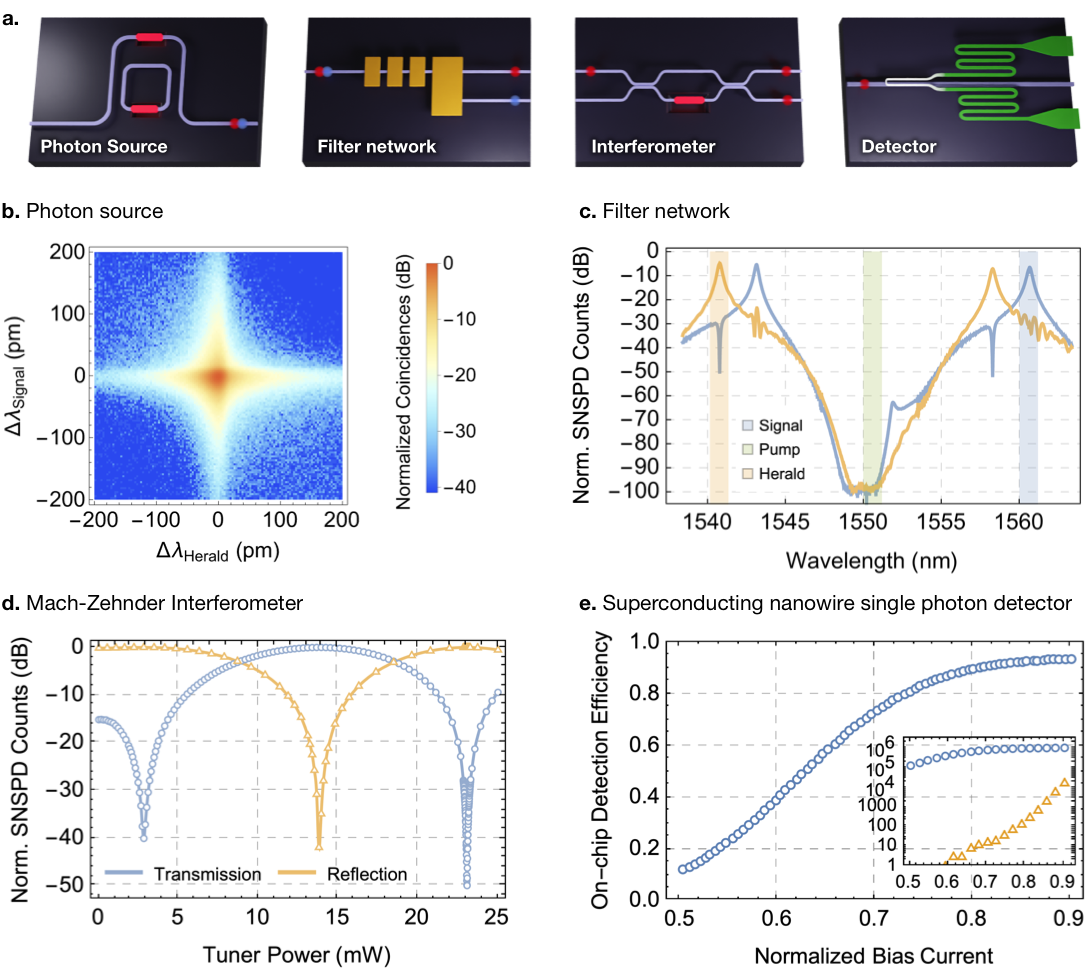}
\caption{\textbf{Key building blocks of the platform.}
\textbf{a}, Schematics of photon source, filter network, interferometer and detector. 
\textbf{b}, Measured joint spectral intensity of an interferometrically-coupled resonator photon source, indicating a spectral purity of $99.5$\% $\pm 0.1$\%. 
\textbf{c}, Response of our pump filter network. We shade the pump, signal and herald frequency bands (shown broader than actual bandwidths \cite{SupMat} for clarity) and show the measured herald (orange) and signal (blue) filter spectrum, characterized with on-chip SNSPDs. 
\textbf{d}, Measured response of a Mach-Zehnder interferometer to heralded single photon illumination in a fully-integrated platform. A dense power sweep around an interference minimum reveals the $>\!50$ dB interference visibility.
\textbf{e}, Measured on-chip detection efficiency as a function of detector bias current ($I_B$) normalized by the detector switching current ($I_\textrm{SW}$), and (inset bottom right) the detector count rate (blue) and dark count rate (orange), per second~\cite{SupMat}. 
}
\label{fig:2}
\end{figure}

Using this stack we build quantum photonic integrated circuits using standard silicon photonic waveguide components, including directional couplers, crossings, and thermal phase shifters. 
We combine these components to produce key building blocks: high-fidelity spontaneous photon pair sources; interferometers for circuit reconfigurability, qubit manipulation, and filtering; and waveguide-integrated single-photon detectors (Fig.~\ref{fig:2}a). We now outline the performance of each of these building blocks. 

\textbf{Photon sources.}
In order to construct entangled resource states and, in turn, an error-correcting code, photonic quantum computers consume many single photons, which must be generated with high efficiency, well-defined timing, and a high repetition rate, whilst also being spectrally pure and indistinguishable \cite{bartolucci2021}. Our single-photon sources use spontaneous four-wave mixing (SFWM) \cite{Silverstone2016} driven by a pulsed laser pump, where the generation of a single photon is probabilistic but heralded by the detection of its pair --- making a heralded single photon source (HSPS).

The visibility of two-photon quantum interference, a key operation in photonic quantum computation, is limited by the spectral purity of the heralding single photons, which is determined by the joint spectral intensity of the photon pairs. We use resonator-based optical waveguide structures to tailor the spectral properties of the photon sources to achieve high spectral purity. The pump is aligned to a resonance frequency and single photons are generated at resonant frequencies spaced symmetrically around it, as illustrated with shaded bands in Fig.~\ref{fig:2}c. Single-ring resonator sources are intrinsically limited to a heralded photon purity of $\sim 93$\% \cite{Vernon17}. We circumvent the spectral purity limitation of single resonator sources using interferometrically-coupled resonator designs \cite{SupMat}, which we characterized to have a measured spectral purity of $99.5$\% $\pm 0.1$\% without spectral filtering (Fig.~\ref{fig:2}b).
 
\textbf{Photon detection.}
Photonic quantum computing relies on heralding the creation of quantum states by detection of correlated photons. Examples include single photon heralding from pair sources, heralded probabilistic resource state generation, and fusion measurements. For fault-tolerance, these functions require near-unit-efficiency single-photon detection. We introduced a niobium nitride (NbN) layer into our photonic stack to enable high-performance manufacturable superconducting nanowire single-photon detectors (SNSPDs) \cite{Goltsman2001,Nam2013}. 

We use a hairpin-shaped SNSPD design \cite{Hu2009}, as depicted in Fig.~\ref{fig:2}a, with a film thickness of $\sim\!5$ nm, nanowire width of $\sim\!90$ nm and detector length of $\geq\!80$ $\mu$m. When operated at $\sim\!2$ K temperature, these detectors exhibit clear plateaus in the photon count rate versus bias current (Fig.~\ref{fig:2}e), indicating high internal detection efficiency. The on-chip detection efficiency is measured via cryogenic electro-optical measurements of waveguide-integrated SNSPDs \cite{SupMat}. Testing of screened SNSPDs yielded a median on-chip efficiency of $93.4\%$ and average value of $88.9\%\pm 3.1\%$ \cite{SupMat}, limited by the hairpin design of the detector. 

\begin{figure*}[t]
\includegraphics[width=\textwidth]{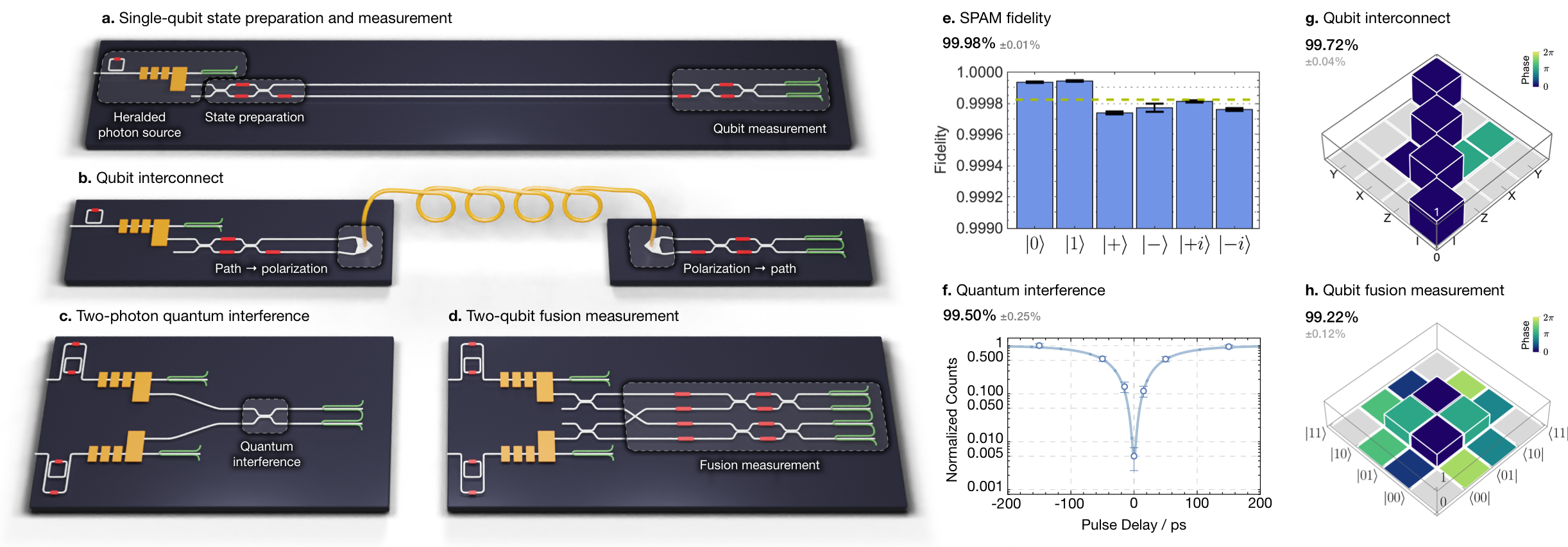}
\caption{\textbf{Quantum Benchmarking Circuits.} These circuits are reconfigurable via thermal phase shifters indicated in red in the schematics. Schematics of: \textbf{a}, quantum state preparation and measurement; \textbf{b}, point-to-point qubit network; \textbf{c}, two-photon quantum (HOM) interference; \textbf{d}, two-qubit fusion measurement. \textbf{e}, SPAM fidelity of the reconstructed state with the target state for Pauli eigenstates. \textbf{f}, HOM interference. \textbf{g}, measured Pauli transfer matrix \cite{Chow2015} of chip-to-chip qubit interconnect channel. \textbf{h}, reconstructed two-qubit density matrix after fusion (grey bars indicate magnitude below 0.01 threshold).}
\label{fig:3}
\end{figure*}

\textbf{Interferometers and filters.} 
Interferometers are a key building block of integrated photonic quantum computing, enabling qubit state preparation and projection, pump filtering, switching networks, resource state generation, and fusion measurements. We use combinations of directional couplers, crossings, and rings to construct ring resonators and Mach-Zehnder-type interferometers. These components have been optimized through design-test cycles and provide predictable performance guaranteed by strict fabrication process control. An example high-contrast Mach-Zehnder interference fringe, measured with co-integrated HSPS and SNSPDs, is shown in Fig.~\ref{fig:2}d, with a $>\!50$ dB extinction ratio. 

Such passive circuits are reconfigurable using thermal phase shifters, which are commonplace in silicon photonics. However, in our platform, the circuit reconfigurability must be compatible with the low temperature operation of integrated superconducting single-photon detectors. Therefore, thermal insulation using undercut regions etched from the silicon substrate (Fig.~\ref{fig:1}a,b,i) is critical.

\section{Integrated heralded single photon generation \& quantum benchmarking circuits}

To date, photonic quantum computing platforms have depended on off-chip single photon sources, off-chip single photon detectors, or both. While sufficient for demonstration purposes, it is very challenging to achieve the heralding efficiency and component density required for practical fault tolerant quantum computing without co-integration of the source, filter and heralding detector. 
Through integration of our key building blocks into our semiconductor platform, we have developed the world's first fully-integrated heralded single photon source --- including source, filtering \& heralding on the same chip. Using this, we construct benchmarking quantum circuits to quantify single-qubit, two-qubit and chip-to-chip qubit interconnect performance, which is summarized in Table \ref{tab:1}.

We selected photonic dies from 300mm wafers using high-volume in-line and end-of-line electric, optical, and electro-optical room-temperature testing; as well as cryogenic electro-optic testing for select parts. For our most complex systems, we package these dies into assemblies (Fig.~\ref{fig:1}d) together with thermal heat-sinks, more than $1000$ electrical connections and up to $200$ optical I/O. We house these packages in cryostats with $\sim\!2$ K base temperature and up to $20$ W cooling capacity (Fig.~\ref{fig:1}k) \cite{SupMat}.

\textbf{Heralded single photon source}.
A high performance HSPS requires engineered SFWM sources, heralding detectors, as well as a high performance filter network on-chip, which we now describe. To separate the bright laser pump from the single photons, we require $\sim\!100$ dB suppression of the pump photons. To achieve this in an integrated circuit, we combine both interferometric in-guide filtering, and shielding of the detectors from out-of-guide scattered pump light. In-guide filtering uses a series of first-order and third-order asymmetric Mach-Zehnder interferometers combined with add-drop resonators to select single source resonances for the herald and the signal photons. Optimizing the free spectral range and coupling values of each element, we achieve pump rejection of $99.1\pm1.2$ dB (Fig.~\ref{fig:2}c) \cite{SupMat}, and the simultaneous rejection of unwanted broadband parametric processes. The signal and herald photons are transmitted through filter networks with approximately $1$ dB of loss. To suppress scattered light, we locally shield the detectors by encasing them in metal (Fig.~\ref{fig:1}a,b,g). The shields are constructed from deep and shallow metal-filled trenches, and back-end-of-line metals. We observe approximately $115$ dB pump power suppression between the pump input and the SNSPDs.

The integrated filters and scattered light shielding, combined with co-integration of SFWMs and SNSPDs allowed for the first demonstration, to our knowledge, of successful on-chip integrated heralded single photon production, with coincidence-to-accidentals (CAR \cite{bienfang2023single}) rates of up to 3000 \cite{SupMat}.

\begin{figure}[ht!]
\includegraphics[width=\columnwidth]{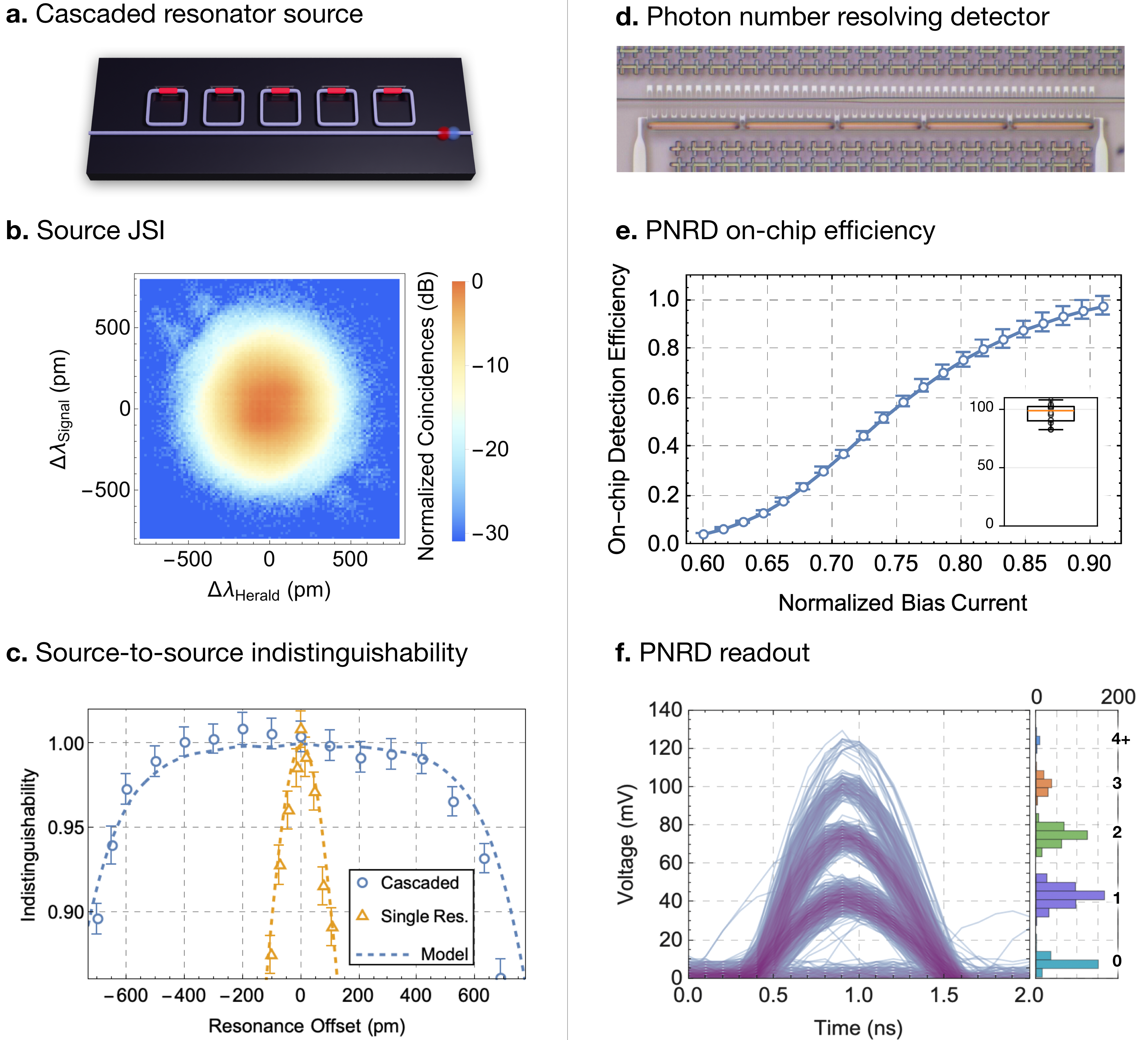}
\caption{\textbf{Cascaded resonator source and PNRD.} 
\textbf{a}, Schematic of the source. 
\textbf{b}, Measured joint spectral intensity of a cascaded resonators source showing up to $99.7$\% purity, assuming flat spectral phase \cite{SupMat}. 
\textbf{c}, measured indistinguishability of two source copies as a function of the resonance wavelength offset \cite{SupMat}.
\textbf{d}, Top-down optical micrograph of a SiN-waveguide-coupled PNRD, where single-photon detectors (SNSPDs) are crossing a waveguide and absorb light from the waveguide through evanescent coupling. Sets of SNSPDs are connected through on-chip resistors to comprise a unit cell. Identical unit cells are connected in series. 
\textbf{e}, On-chip detection efficiency for the PNRD shown in \textbf{d} as a function of normalized bias current, showing the average across 6 unique devices \cite{SupMat}. 
\textbf{e}-inset, distribution of single-shot detection efficiency for each of the unique devices biased at $\sim\!0.9 I_\textrm{sw}$ at 2 input power levels. 
\textbf{f}, (left) Persistent plot of the electrical photodetection signal (voltage traces) of a 4-unit-cell PNRD. The traces were recorded using a cryogenic amplifier. The voltage traces show 5 distinct levels, corresponding to 0, 1, 2, 3 and 4 unit cells detecting photons simultaneously. (To right, voltage trace histogram).}
\label{fig:4}
\end{figure}

\textbf{Single qubit state preparation and measurement (SPAM).} We prepare a path encoded qubit \cite{Politi2008} using a heralded photon and two-mode interferometers, as illustrated in Fig.~\ref{fig:3}a. We measure the path encoded qubit using a two-mode interferometer and SNSPDs. 
The state of the single photon in two optical modes is controlled by two thermal phase shifters, which enable the encoding of arbitrary qubit states. 
We observe an average single-qubit preparation and measurement fidelity of $99.98\% \pm 0.01\%$ (Fig.~\ref{fig:3}e), conditional on the photon being detected \cite{SupMat}. 
Aiming to separate the impact of the HSPS's signal to noise ratio (SNR), we repeat the measurement on a different but equivalent chip, using bright coherent light and off-chip photodetectors, achieving a fidelity of $99.996\%\pm0.003\%$ \cite{SupMat}, showing that higher SPAM fidelity will be possible with improved HSPS SNR.

\textbf{Chip-to-chip qubit interconnect.} Networking of quantum modules has seen growing interest as various technologies seek to scale beyond the boundary of a single chip, trap or reticle. Telecom-wavelength photonic qubits are naturally suited for transmission through optical fiber, without the need for quantum transduction \cite{Lauk2020}. Additionally, optical fiber-based networking can enable additional novel functionality such as interleaving \cite{Interleaving2021} and active-volume compilation \cite{litinski2022active} leading to large resource savings for fault-tolerant algorithms. To demonstrate the networking capability of our photonic qubits we build a point-to-point qubit network (Fig.~\ref{fig:3}b) and assess the fidelity of qubits after propagating between modules. We prepare high-fidelity single qubit states using the same qubit state preparation circuit as described above, and convert to polarization encoding using a two-dimensional grating coupler-based path-to-polarization converter \cite{Llewellyn2020}. We transmit the qubit over 42m of standard telecommunications grade optical fiber, before converting to path encoding at the receiving module and performing on-chip qubit state measurement. The transmission and receiving modules both use on-chip superconducting detectors, and operate at liquid helium temperature. We determined the Pauli transfer matrix \cite{Chow2015} fidelity between the physical channel and the identity operation, conditional on photon arrival, to be $99.72$\% $\pm 0.04$\% (Fig.~\ref{fig:3}g) \cite{SupMat}. The system exhibits high loss associated with fiber-to-chip coupling by grating couplers, which will be overcome in future systems using edge-coupled devices (discussed below). 

\textbf{Two-photon quantum interference.}
To benchmark our integrated single photon sources, we measure Hong-Ou-Mandel (HOM) quantum interference between heralded photons from two independent sources integrated on the same chip (Fig.~\ref{fig:3}c). The measured visibility depends on many factors including indistinguishability, spectral purity, number purity, signal-to-noise ratio, and system detection efficiency. To control these, we implement a single system that integrates the technologies described above: high purity, tunable photon pair sources; high extinction filter network; and high efficiency and shielded SNSPDs.

The on-chip HOM quantum interference between heralded photons from different sources, without significant spectral filtering, was $99.50\%\pm0.25\%$ (Fig.~\ref{fig:3}f), which to our knowledge is the highest measured in any platform. The experiment was performed at a pump repetition rate of 125 MHz, with a source CAR of $929\pm4$, a heralded $g^{(2)}(0)=0.00358\pm0.00024$, and a maximum Klyshko efficiency of $\sim\!26$\% \cite{SupMat}.

\textbf{Two-qubit fusion.} 
Bell fusion is a projective measurement onto two-qubit Bell states and is the prototypical example of the class of measurements which underpins the FBQC paradigm \cite{FBQC2023}. We implement Bell fusion using Type II fusion measurements \cite{Browne2004} on dual rail qubits. Type II fusion uses a four mode linear optical circuit followed by photon detection. It requires both single-qubit interference and interference between qubits, enabled by high performance qubit preparations and high-visibility two-photon quantum interference, respectively.

We demonstrate that the fusion operation can perform a high fidelity projection onto a Bell state, using the benchmarking circuit in (Fig.~\ref{fig:3}d). Two independent path-encoded single qubits are prepared in the product state $\ket{+-}$. Using a reconfigurable fusion-measurement network, paths are then exchanged between the qubits and the resulting state is measured via single-qubit measurements. When a photon is detected in each pair of detectors, we measure a fidelity of $99.22\%\pm0.12\%$ with the ideal Bell state. The density matrix is shown in Fig.~\ref{fig:3}h.

\begin{table}[]
\resizebox{\columnwidth}{!}{%
\begin{tabular}{@{}ccc@{}}
\toprule
                              & Metric                & Experiment Value (\%)                                                                                                                \\ \midrule
\multirow{2}{*}{Single-Qubit} & SPAM fidelity         & \begin{tabular}[c]{@{}c@{}}$99.98 \pm 0.01$ \\ ~~$99.996 \pm 0.003^*$\end{tabular} \\ \cmidrule(l){2-3} 
                              & Chip-to-chip fidelity & $99.72 \pm 0.04$                                                                                                                       \\ \midrule
\multirow{2}{*}{Two-Qubit}    & Quantum interference visibility        & $99.50 \pm 0.25$                                                                                                                       \\ \cmidrule(l){2-3} 
                              & Bell fidelity         & $99.22 \pm 0.12$                                                                                                                           \\ \bottomrule
\end{tabular}%
}
\caption{Single- and two-qubit performance metrics. Not accounting for loss.
*Second SPAM fidelity listed above is measured with bright light and off-chip detectors, see main text. }
\label{tab:1}
\end{table}

\section{Next generation technologies}
The performance of the baseline technology described above is still not sufficient for useful photonic quantum computing. In particular, silicon waveguides incur too much propagation loss for fault tolerance, photon sources require complex and power-hungry tuning, and high-speed optical switching is unavoidably necessary to overcome the intrinsic non-determinism of the spontaneous single photon sources.

We now describe some of the critical developments towards higher performance and additional functionality in our next-generation technology platforms, derived from multiple process flows. We focus on advanced photon sources, high-efficiency photon-number-resolving detection, low-loss waveguides, high-efficiency fiber-to-chip coupling, and on-chip electro-optic phase shifters. 

\textbf{Cascaded resonator source.} 
The key performance metrics for photon sources are two-photon interference visibility and photon efficiency. However, there are additional characteristics that must be addressed to enable the operation of devices at the scale of useful quantum computers. Two important considerations are the pump power required to drive the SFWM process, and the thermal-power dissipated at cryogenic temperatures to control and tune the source. We have implemented a cascaded resonator source which addresses these aspects simultaneously, enabling high purity and spectral indistinguishability, reduced pump power, and increased tolerance to fabrication imperfections.

The source comprises multiple integrated resonators coupled to a single bus waveguide (Fig.~\ref{fig:4}a). Through joint optimization of the resonator-bus coupling, the resonance wavelengths, and the pump spectral amplitude, the joint spectral intensity of the source can be engineered. Our 24-resonator device has a measured upper bounded purity of up to $99.7$\%, assuming flat spectral phase (Fig.~\ref{fig:4}b), whilst using an order of magnitude less pump power than the interferometically-coupled source design.

This cascaded resonators source addresses indistinguishability in a novel way. The spectrum of the photon pairs is fixed by the pump wavelength and not by the resonant wavelength of the device. Thus, global resonance shifts (e.g. from fabrication variations) have minimal impact on the spectral indistinguishability of photons generated from different devices. Fig.~\ref{fig:4}c, shows the measured indistinguishability between two sources as a function of resonance shift. Using thermal tuners, we aligned two devices to the optimal operating point and applied a controlled global resonance shift to one cascaded resonator source, to simulate the impact of fabrication variation. In this implementation, we achieve $>99\%$ two-source indistinguishability over a $\pm400$ pm resonance shift window, compared to less than $\pm40$ pm for a single-ring source. 

The built-in tolerance of the cascaded resonator source to device-to-device global wavelength variations, together with state-of-the-art foundry process and fabrication control, can enable robust, tunerless and manufacturable indistinguishable photon sources.

\textbf{Photon-number-resolving detectors (PNRDs).} 
The waveguide-integrated manufacturable single-photon detectors presented earlier, while transformative, lack the photon-number-resolving capability required for FBQC. The ability to distinguish low photon numbers in detection, and to herald on that information, allows for both the removal of higher-order photon number states generated in SFWM sources, and the identification of unwanted events in fusion-based entangled state generation and computation \cite{bartolucci2021}. 

Spatial multiplexing \cite{Jahanmirinejad:12} of many SNSPD-like detector elements, as shown in Fig.~\ref{fig:4}d, can be used to assemble a scalable detector with effective photon-number resolution. In these PNRDs, the number of detected photons is approximately proportional to the amplitude of the detector output voltage. To validate this concept, we have produced waveguide-integrated PNRDs with 4 and 5 unit cells, with the best performing designs yielding on-chip detection efficiencies of $98.9\%$ (median) and $96.3\pm 3.9\%$ (mean) (Fig.~\ref{fig:4}e). These detectors have the ability to resolve $0, 1, 2, 3, 4+$ photons, as shown in the histogram of Fig.~\ref{fig:4}f.

\begin{figure}[t]
\includegraphics[width=\columnwidth]{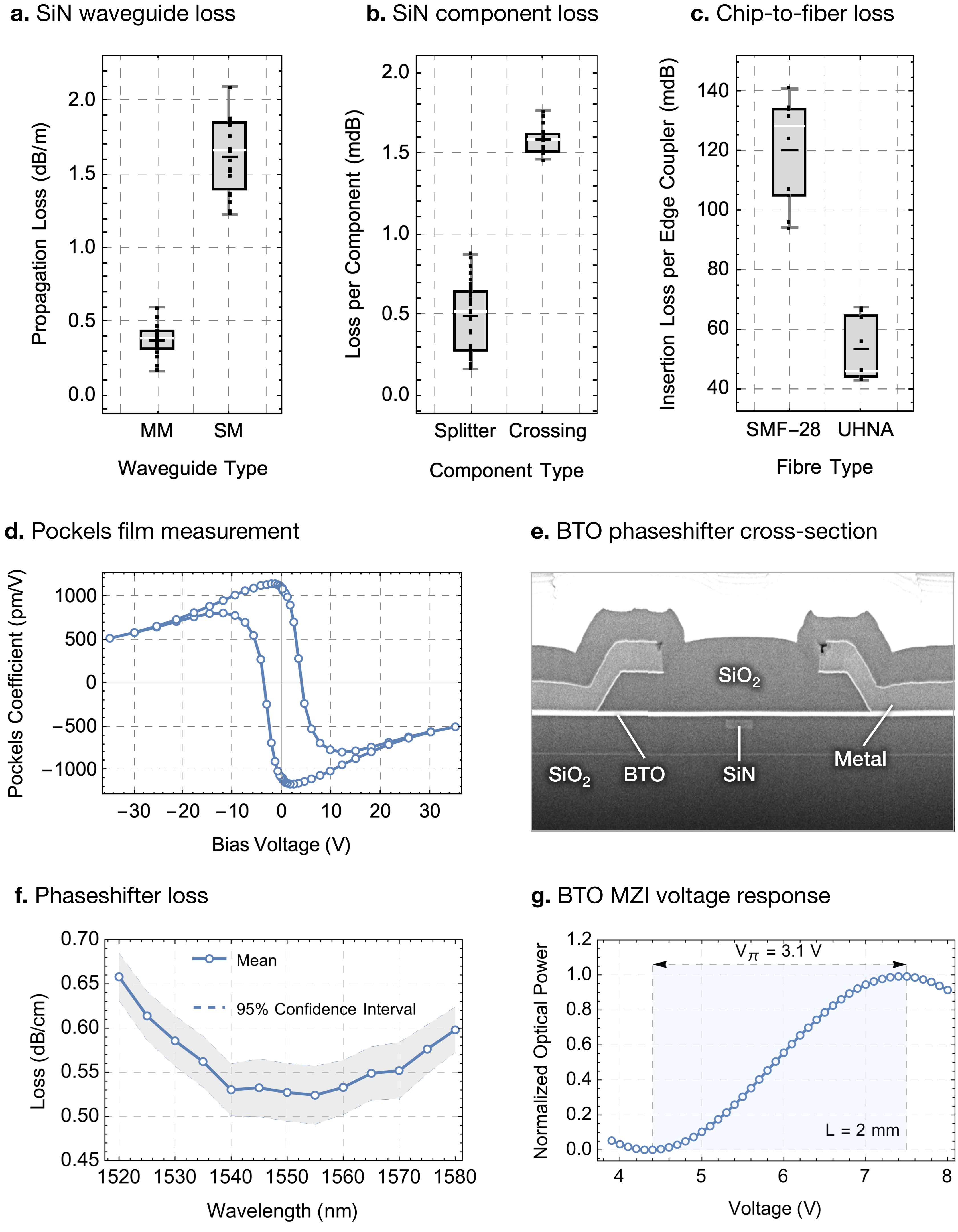}
\caption{\textbf{Waveguide and component loss, and BTO optical switch. a,b,c,} Loss of SiN-based components with mean (black) and median (white). \textbf{a}, SiN waveguide loss measurement, showing results across example wafers for both multi-mode (MM) and single-mode (SM) waveguides \cite{SupMat}. \textbf{b}, SiN component loss for waveguide splitters and crossings \cite{SupMat}. \textbf{c}, Chip-to-fibre loss. The fiber-to-chip coupling is measured in the low-loss regime using repeated transmission measurements on two exemplary devices designed for SMF-28 fiber and an exemplary device designed for UHNA fiber \cite{SupMat}. \textbf{d}, Free-space electro-optic measurement of the effective Pockels coefficient of an MBE-grown BTO film, with hysteresis. \textbf{e}, Scanning electron microscope cross-section of a fully fabricated BTO-on-SiN phase shifter. \textbf{f}, Cutback-based propagation loss measurement of a BTO-on-SiN phase shifter (data points and guide-line), with 95\% confidence intervals provided (dashed lines). \textbf{g}, measured optical transmission of a Mach-Zehnder interferometer (MZI) with a $L=2$ mm-long BTO phase shifter. A voltage was applied to one arm of the MZI, resulting in a $V_\pi L = 0.62$ V.cm in a non-push-pull configuration \cite{SupMat}, where $V_\pi$ is the voltage required to change the phase by $\pi$ radians.}
\label{fig:5}
\end{figure}

\textbf{Low-loss silicon nitride waveguides, directional couplers and crossings.} 
Silicon-on-insulator waveguides are limited in waveguide propagation loss due to their large refractive index contrast \cite{sacher2018}. Silicon nitride (SiN) waveguides, on the other hand, have lower refractive index contrast, offering a good compromise between confinement and sensitivity to manufacturing variations \cite{sacher2018}. We have demonstrated single-mode SiN waveguide loss of $1.58\pm 0.50$ dB/m and multimode waveguide loss of $0.39\pm 0.12$ dB/m (Fig.~\ref{fig:5}a). In this same platform, we have implemented waveguide crossings with $1.6\pm0.1$ mdB loss, and waveguide splitters with $0.5\pm0.2$ mdB loss (Fig.~\ref{fig:5}b). 

SiN also provides advantages for photon generation. The ultra-low-loss combined with its Kerr non-linearity supports SFWM with high signal-to-noise ratio. Further, there is an absence of non-linear loss, allowing sources to operate with low loss at high-pair rates, unlike silicon, where two-photon absorption degrades performance \cite{Bristow2007,Rosenfeld2020}. 

\textbf{Fiber-to-chip coupling.} 
Low-loss coupling of light from optical fibers to our quantum photonic chips is required to make fiber networking practical. We implement novel edge coupler designs which minimize mode overlap and mode conversion loss, enabling high-performance fiber-to-chip coupling. 
A key challenge is to convert the highly confined on-chip waveguide mode to match the much larger mode of optical fiber. To measure the insertion loss of the edge coupler, a chip is positioned between input and output optical fibers using high-precision optical alignment stages. Figure \ref{fig:5}c shows repeated measurements from two of our best chip-to-fiber coupler designs, with coupling loss to standard telecommunications grade optical fiber (SMF-28) of $120\pm17$ mdB, and coupling loss to high-numerical aperture fiber (UHNA4) of $53\pm12$ mdB. 

\textbf{Electro-optic switching.} 
In order to overcome the intrinsic non-determinism of both spontaneous sources and fusion gates, photonic quantum computing will require beyond-state-of-the-art high-speed optical switches, in order to enable large optical networks which can be rapidly reconfigured based on the results of previous heralded photon generation, entangling gates, and fusion outcomes \cite{SwitchNetworks2021}. 
The key component required for such switching networks is a high-speed, low-loss electro-optic phase shifter. Complex N$\times$M networks may be constructed by embedding this phase shifter into passive interferometers constructed from the beamsplitter and crossing devices previously described \cite{SwitchNetworks2021}.

The performance of the phaseshifter is fundamentally constrained by the choice of electo-optic material. We incorporate BTO \cite{Eltes2020} into our photonic stack as the electro-optic phase shifter. We have developed a proprietary process for the growth of high-quality BTO films using molecular beam epitaxy, compatible with foundry processes, on full 300mm silicon wafers. We achieved a $3\sigma$ thickness uniformity of $<3\%$ across the entire 300mm wafer, with electro-optic Pockels values of $>1000$ pm/V (compared to $\sim\!30$ pm/V for lithium niobate \cite{Zhu21}), measured through free-space Pockels measurements (Fig.~\ref{fig:5}d). 

Our fabricated $2\times2$ BTO Mach-Zehnder switches include a 2mm-long phase shifter section, with a propagation loss of $53\pm3$ dB/m and a DC $V_\pi L$ of $0.62$ V.cm. This gives a phase shifter insertion loss of $\sim100$ mdB and a phase shifter half-wave voltage-loss product ($V_\pi L\alpha$) of $0.33\pm0.02$ V.dB (Fig.~\ref{fig:5}d,f,g), enabling a path to construction of larger N$\times$M low-loss switching networks required for photonic quantum computing. 

\section{Conclusion}
We have described modifications made to an industrial semiconductor manufacturing process for integrated quantum photonics, demonstrating record performance. Through the addition of new materials, designs and process steps, we have enabled volume manufacturing of heralded photon sources and superconducting single photon detectors, together with photon manipulation via interferometry, tunability, and control of unwanted light. We have also described higher-performing devices, towards a resolution of the outstanding limitations of this baseline platform. 

Fusion-based quantum computing supports fault-tolerant protocols which can tolerate optical loss of order $10$\% during a photon lifetime, with per-qubit errors in the fusion network of order $1$\% \cite{bombin2023increasing,FBQC2023,bombin2023_2,MIP}. 
Here we have demonstrated a feature-complete set of optical components for FBQC, each with optical losses at the few-percent or below level, as well as packaged devices demonstrating high-visibility interference, distribution and measurement functionalities of photonic qubits, all with sub-percent error levels.

Improvements to the platform and processes are still required. It will be necessary to further reduce SiN materials and component losses, improve filter performance, and increase detector efficiency to push overall photon loss and fidelity. Some specific examples of the remaining challenges are: implementation of low-loss N$\times$M fast switches towards a multiplexed photon source; repeatable alignment and packaging of ultra-low-loss chip-to-fiber edge connects; and improved targeting and robustness of photonic designs to minimize the need for tuning and trimming with heaters, thus further reducing the heat load at cryogenic temperatures. 

Finally, we note that the platforms we have developed, and their future improvements, are highly flexible. Component arrangements are highly configurable, making the system suitable for different variations of quantum computer architectures, different quantum technology applications, and indeed other photonic technologies. The ability to connect chips by fiber with very low loss makes the system technologically scalable across large numbers of photonic dies, and allows for future networking or connections between different systems in a range of application spaces. Although the singular intent of our development is a useful fault-tolerant quantum computer, we hope the impact of our industrially-manufacturable quantum photonic platform will be broad and substantial.

\begin{flushleft}

\noindent\hrulefill

\textbf{PsiQuantum Team.}
\newline
\newline

\textbf{Koen Alexander$^{1}$, Andrea Bahgat$^{1}$, Avishai Benyamini$^{1}$, Dylan Black$^{1}$, Damien Bonneau$^{1}$, Stanley Burgos$^{1}$, Ben Burridge$^{1,2}$, Geoff Campbell$^{1}$, Gabriel Catalano$^{1}$, Alex Ceballos$^{1}$, Chia-Ming Chang$^{1}$, CJ Chung$^{1}$, Fariba Danesh$^{1}$, Tom Dauer$^{1}$, Michael Davis$^{1}$, Eric Dudley$^{1}$, Ping Er-Xuan$^{1}$, Josep Fargas$^{1}$, Alessandro Farsi$^{1}$, Colleen Fenrich$^{1}$, Jonathan Frazer$^{1,2}$, Masaya Fukami$^{1}$, Yogeeswaran Ganesan$^{1}$, Gary Gibson$^{1}$, Mercedes Gimeno-Segovia$^{1}$, Sebastian Goeldi$^{1}$, Patrick Goley$^{1}$, Ryan Haislmaier$^{1}$, Sami Halimi$^{1}$, Paul Hansen$^{1}$, Sam Hardy$^{1}$, Jason Horng$^{1}$, Matthew House$^{1}$, Hong Hu$^{1}$, Mehdi Jadidi$^{1}$, Henrik Johansson$^{1}$, Thomas Jones$^{1,2}$, Vimal Kamineni$^{1}$, Nicholas Kelez$^{1}$, Ravi Koustuban$^{1}$, George Kovall$^{1}$, Peter Krogen$^{1}$, Nikhil Kumar$^{1}$, Yong Liang$^{1}$, Nicholas LiCausi$^{1}$, Dan Llewellyn$^{1,2}$, Kimberly Lokovic$^{1}$, Michael Lovelady$^{1}$, Vitor Manfrinato$^{1}$, Ann Melnichuk$^{1}$, Mario Souza$^{1}$, Gabriel Mendoza$^{1}$, Brad Moores$^{1}$, Shaunak Mukherjee$^{1}$, Joseph Munns$^{1}$, Francois-Xavier Musalem$^{1}$, Faraz Najafi$^{1}$, Jeremy L. O'Brien$^{1}$, J. Elliott Ortmann$^{1}$, Sunil Pai$^{1}$, Bryan Park$^{1}$, Hsuan-Tung Peng$^{1}$, Nicholas Penthorn$^{1}$, Brennan Peterson$^{1}$, Matt Poush$^{1}$, Geoff J. Pryde$^{1}$, Tarun Ramprasad$^{1,2}$, Gareth Ray$^{1,2}$, Angelita Rodriguez$^{1}$, Brian Roxworthy$^{1}$, Terry Rudolph$^{1}$, Dylan J. Saunders$^{1}$, Pete Shadbolt$^{1}$, Deesha Shah$^{1}$, Hyungki Shin$^{1}$, Jake Smith$^{1}$, Ben Sohn$^{1}$, Young-Ik Sohn$^{1}$, Gyeongho Son$^{1}$, Chris Sparrow$^{1}$, Matteo Staffaroni$^{1}$, Camille Stavrakas$^{1}$, Vijay Sukumaran$^{1}$, Davide Tamborini$^{1}$, Mark G. Thompson$^{1,2,}$\textsuperscript{\Letter}, Khanh Tran$^{1}$, Mark Triplet$^{1}$, Maryann Tung$^{1}$, Alexey Vert$^{1}$, Mihai D. Vidrighin$^{1,2}$, Ilya Vorobeichik$^{1}$, Peter Weigel$^{1}$, Mathhew Wingert$^{1}$, Jamie Wooding$^{1}$, Xinran Zhou$^{1}$}
\newline
\newline
$^{1}$PsiQuantum Corp., Palo Alto 94304 CA, USA \\
$^{2}$PsiQuantum Ltd., Daresbury WA4 4FS, UK \\
\textsuperscript{\Letter}e-mail: mark@psiquantum.com
\end{flushleft}


\begin{thebibliography}{38}%
\makeatletter
\providecommand \@ifxundefined [1]{%
 \@ifx{#1\undefined}
}%
\providecommand \@ifnum [1]{%
 \ifnum #1\expandafter \@firstoftwo
 \else \expandafter \@secondoftwo
 \fi
}%
\providecommand \@ifx [1]{%
 \ifx #1\expandafter \@firstoftwo
 \else \expandafter \@secondoftwo
 \fi
}%
\providecommand \natexlab [1]{#1}%
\providecommand \enquote  [1]{``#1''}%
\providecommand \bibnamefont  [1]{#1}%
\providecommand \bibfnamefont [1]{#1}%
\providecommand \citenamefont [1]{#1}%
\providecommand \href@noop [0]{\@secondoftwo}%
\providecommand \href [0]{\begingroup \@sanitize@url \@href}%
\providecommand \@href[1]{\@@startlink{#1}\@@href}%
\providecommand \@@href[1]{\endgroup#1\@@endlink}%
\providecommand \@sanitize@url [0]{\catcode `\\12\catcode `\$12\catcode
  `\&12\catcode `\#12\catcode `\^12\catcode `\_12\catcode `\%12\relax}%
\providecommand \@@startlink[1]{}%
\providecommand \@@endlink[0]{}%
\providecommand \url  [0]{\begingroup\@sanitize@url \@url }%
\providecommand \@url [1]{\endgroup\@href {#1}{\urlprefix }}%
\providecommand \urlprefix  [0]{URL }%
\providecommand \Eprint [0]{\href }%
\providecommand \doibase [0]{http://dx.doi.org/}%
\providecommand \selectlanguage [0]{\@gobble}%
\providecommand \bibinfo  [0]{\@secondoftwo}%
\providecommand \bibfield  [0]{\@secondoftwo}%
\providecommand \translation [1]{[#1]}%
\providecommand \BibitemOpen [0]{}%
\providecommand \bibitemStop [0]{}%
\providecommand \bibitemNoStop [0]{.\EOS\space}%
\providecommand \EOS [0]{\spacefactor3000\relax}%
\providecommand \BibitemShut  [1]{\csname bibitem#1\endcsname}%
\let\auto@bib@innerbib\@empty
%</preamble>
\bibitem [{\citenamefont {Fowler}\ \emph {et~al.}(2004)\citenamefont {Fowler},
  \citenamefont {Devitt},\ and\ \citenamefont {Hollenberg}}]{Hollenberg2004}%
  \BibitemOpen
  \bibfield  {author} {\bibinfo {author} {\bibfnamefont {A.~G.}\ \bibnamefont
  {Fowler}}, \bibinfo {author} {\bibfnamefont {S.~J.}\ \bibnamefont {Devitt}},
  \ and\ \bibinfo {author} {\bibfnamefont {L.~C.~L.}\ \bibnamefont
  {Hollenberg}},\ }\href@noop {} {\bibfield  {journal} {\bibinfo  {journal}
  {Quantum Info. Comput.}\ }\textbf {\bibinfo {volume} {4}},\ \bibinfo {pages}
  {237–} (\bibinfo {year} {2004})}\BibitemShut {NoStop}%
\bibitem [{\citenamefont {O'Brien}(2007)}]{OBrien2007}%
  \BibitemOpen
  \bibfield  {author} {\bibinfo {author} {\bibfnamefont {J.~L.}\ \bibnamefont
  {O'Brien}},\ }\href {\doibase 10.1126/science.1142892} {\bibfield  {journal}
  {\bibinfo  {journal} {Science}\ }\textbf {\bibinfo {volume} {318}},\ \bibinfo
  {pages} {1567} (\bibinfo {year} {2007})}\BibitemShut {NoStop}%
\bibitem [{\citenamefont {Knill}\ \emph {et~al.}(2001)\citenamefont {Knill},
  \citenamefont {Laflamme},\ and\ \citenamefont {Milburn}}]{KLM}%
  \BibitemOpen
  \bibfield  {author} {\bibinfo {author} {\bibfnamefont {E.}~\bibnamefont
  {Knill}}, \bibinfo {author} {\bibfnamefont {R.}~\bibnamefont {Laflamme}}, \
  and\ \bibinfo {author} {\bibfnamefont {G.~J.}\ \bibnamefont {Milburn}},\
  }\href {\doibase 10.1038/35051009} {\bibfield  {journal} {\bibinfo  {journal}
  {Nature}\ }\textbf {\bibinfo {volume} {409}},\ \bibinfo {pages} {46}
  (\bibinfo {year} {2001})}\BibitemShut {NoStop}%
\bibitem [{\citenamefont {Ralph}\ \emph {et~al.}(2001)\citenamefont {Ralph},
  \citenamefont {White}, \citenamefont {Munro},\ and\ \citenamefont
  {Milburn}}]{Ralph2001}%
  \BibitemOpen
  \bibfield  {author} {\bibinfo {author} {\bibfnamefont {T.~C.}\ \bibnamefont
  {Ralph}}, \bibinfo {author} {\bibfnamefont {A.~G.}\ \bibnamefont {White}},
  \bibinfo {author} {\bibfnamefont {W.~J.}\ \bibnamefont {Munro}}, \ and\
  \bibinfo {author} {\bibfnamefont {G.~J.}\ \bibnamefont {Milburn}},\ }\href
  {\doibase 10.1103/PhysRevA.65.012314} {\bibfield  {journal} {\bibinfo
  {journal} {Phys. Rev. A}\ }\textbf {\bibinfo {volume} {65}},\ \bibinfo
  {pages} {012314} (\bibinfo {year} {2001})}\BibitemShut {NoStop}%
\bibitem [{\citenamefont {Yoran}\ and\ \citenamefont
  {Reznik}(2003)}]{Yoran2003}%
  \BibitemOpen
  \bibfield  {author} {\bibinfo {author} {\bibfnamefont {N.}~\bibnamefont
  {Yoran}}\ and\ \bibinfo {author} {\bibfnamefont {B.}~\bibnamefont {Reznik}},\
  }\href {\doibase 10.1103/PhysRevLett.91.037903} {\bibfield  {journal}
  {\bibinfo  {journal} {Phys. Rev. Lett.}\ }\textbf {\bibinfo {volume} {91}},\
  \bibinfo {pages} {037903} (\bibinfo {year} {2003})}\BibitemShut {NoStop}%
\bibitem [{\citenamefont {Nielsen}(2004)}]{Nielsen2004}%
  \BibitemOpen
  \bibfield  {author} {\bibinfo {author} {\bibfnamefont {M.~A.}\ \bibnamefont
  {Nielsen}},\ }\href {\doibase 10.1103/PhysRevLett.93.040503} {\bibfield
  {journal} {\bibinfo  {journal} {Phys. Rev. Lett.}\ }\textbf {\bibinfo
  {volume} {93}},\ \bibinfo {pages} {040503} (\bibinfo {year}
  {2004})}\BibitemShut {NoStop}%
\bibitem [{\citenamefont {Browne}\ and\ \citenamefont
  {Rudolph}(2005)}]{Browne2004}%
  \BibitemOpen
  \bibfield  {author} {\bibinfo {author} {\bibfnamefont {D.~E.}\ \bibnamefont
  {Browne}}\ and\ \bibinfo {author} {\bibfnamefont {T.}~\bibnamefont
  {Rudolph}},\ }\href {\doibase 10.1103/PhysRevLett.95.010501} {\bibfield
  {journal} {\bibinfo  {journal} {Phys. Rev. Lett.}\ }\textbf {\bibinfo
  {volume} {95}},\ \bibinfo {pages} {010501} (\bibinfo {year}
  {2005})}\BibitemShut {NoStop}%
\bibitem [{\citenamefont {Li}\ \emph {et~al.}(2015)\citenamefont {Li},
  \citenamefont {Humphreys}, \citenamefont {Mendoza},\ and\ \citenamefont
  {Benjamin}}]{Ying2015}%
  \BibitemOpen
  \bibfield  {author} {\bibinfo {author} {\bibfnamefont {Y.}~\bibnamefont
  {Li}}, \bibinfo {author} {\bibfnamefont {P.~C.}\ \bibnamefont {Humphreys}},
  \bibinfo {author} {\bibfnamefont {G.~J.}\ \bibnamefont {Mendoza}}, \ and\
  \bibinfo {author} {\bibfnamefont {S.~C.}\ \bibnamefont {Benjamin}},\ }\href
  {\doibase 10.1103/PhysRevX.5.041007} {\bibfield  {journal} {\bibinfo
  {journal} {Phys. Rev. X}\ }\textbf {\bibinfo {volume} {5}},\ \bibinfo {pages}
  {041007} (\bibinfo {year} {2015})}\BibitemShut {NoStop}%
\bibitem [{\citenamefont {Rudolph}(2017)}]{Rudolph2017}%
  \BibitemOpen
  \bibfield  {author} {\bibinfo {author} {\bibfnamefont {T.}~\bibnamefont
  {Rudolph}},\ }\href {\doibase 10.1063/1.4976737} {\bibfield  {journal}
  {\bibinfo  {journal} {APL Photonics}\ }\textbf {\bibinfo {volume} {2}},\
  \bibinfo {pages} {030901} (\bibinfo {year} {2017})}\BibitemShut {NoStop}%
\bibitem [{\citenamefont {Gol’tsman}\ \emph {et~al.}(2001)\citenamefont
  {Gol’tsman}, \citenamefont {Okunev}, \citenamefont {Chulkova},
  \citenamefont {Lipatov}, \citenamefont {Semenov}, \citenamefont {Smirnov},
  \citenamefont {Voronov}, \citenamefont {Dzardanov}, \citenamefont
  {Williams},\ and\ \citenamefont {Sobolewski}}]{Goltsman2001}%
  \BibitemOpen
  \bibfield  {author} {\bibinfo {author} {\bibfnamefont {G.~N.}\ \bibnamefont
  {Gol’tsman}}, \bibinfo {author} {\bibfnamefont {O.}~\bibnamefont {Okunev}},
  \bibinfo {author} {\bibfnamefont {G.}~\bibnamefont {Chulkova}}, \bibinfo
  {author} {\bibfnamefont {A.}~\bibnamefont {Lipatov}}, \bibinfo {author}
  {\bibfnamefont {A.}~\bibnamefont {Semenov}}, \bibinfo {author} {\bibfnamefont
  {K.}~\bibnamefont {Smirnov}}, \bibinfo {author} {\bibfnamefont
  {B.}~\bibnamefont {Voronov}}, \bibinfo {author} {\bibfnamefont
  {A.}~\bibnamefont {Dzardanov}}, \bibinfo {author} {\bibfnamefont
  {C.}~\bibnamefont {Williams}}, \ and\ \bibinfo {author} {\bibfnamefont
  {R.}~\bibnamefont {Sobolewski}},\ }\href {\doibase 10.1063/1.1388868}
  {\bibfield  {journal} {\bibinfo  {journal} {Applied Physics Letters}\
  }\textbf {\bibinfo {volume} {79}},\ \bibinfo {pages} {705} (\bibinfo {year}
  {2001})},\ \Eprint
  {http://arxiv.org/abs/https://pubs.aip.org/aip/apl/article-pdf/79/6/705/18559493/705\_1\_online.pdf}
  {https://pubs.aip.org/aip/apl/article-pdf/79/6/705/18559493/705\_1\_online.pdf}
  \BibitemShut {NoStop}%
\bibitem [{\citenamefont {Marsili}\ \emph {et~al.}(2013)\citenamefont
  {Marsili}, \citenamefont {Verma}, \citenamefont {Stern}, \citenamefont
  {Harrington}, \citenamefont {Lita}, \citenamefont {Gerrits}, \citenamefont
  {Vayshenker}, \citenamefont {Baek}, \citenamefont {Shaw}, \citenamefont
  {Mirin},\ and\ \citenamefont {Nam}}]{Nam2013}%
  \BibitemOpen
  \bibfield  {author} {\bibinfo {author} {\bibfnamefont {F.}~\bibnamefont
  {Marsili}}, \bibinfo {author} {\bibfnamefont {V.~B.}\ \bibnamefont {Verma}},
  \bibinfo {author} {\bibfnamefont {J.~A.}\ \bibnamefont {Stern}}, \bibinfo
  {author} {\bibfnamefont {S.}~\bibnamefont {Harrington}}, \bibinfo {author}
  {\bibfnamefont {A.~E.}\ \bibnamefont {Lita}}, \bibinfo {author}
  {\bibfnamefont {T.}~\bibnamefont {Gerrits}}, \bibinfo {author} {\bibfnamefont
  {I.}~\bibnamefont {Vayshenker}}, \bibinfo {author} {\bibfnamefont
  {B.}~\bibnamefont {Baek}}, \bibinfo {author} {\bibfnamefont {M.~D.}\
  \bibnamefont {Shaw}}, \bibinfo {author} {\bibfnamefont {R.~P.}\ \bibnamefont
  {Mirin}}, \ and\ \bibinfo {author} {\bibfnamefont {S.~W.}\ \bibnamefont
  {Nam}},\ }\href {\doibase 10.1038/nphoton.2013.13} {\bibfield  {journal}
  {\bibinfo  {journal} {Nature Photonics}\ }\textbf {\bibinfo {volume} {7}},\
  \bibinfo {pages} {210} (\bibinfo {year} {2013})}\BibitemShut {NoStop}%
\bibitem [{\citenamefont {Geng}(2018)}]{geng2018semiconductor}%
  \BibitemOpen
  \bibfield  {author} {\bibinfo {author} {\bibfnamefont {H.}~\bibnamefont
  {Geng}},\ }\href@noop {} {\emph {\bibinfo {title} {Semiconductor
  Manufacturing Handbook. 2nd ed}}}\ (\bibinfo  {publisher} {New York:
  McGraw-Hill Education},\ \bibinfo {year} {2018})\BibitemShut {NoStop}%
\bibitem [{\citenamefont {Bartolucci}\ \emph {et~al.}(2023)\citenamefont
  {Bartolucci}, \citenamefont {Birchall}, \citenamefont {Bomb{\'\i}n},
  \citenamefont {Cable}, \citenamefont {Dawson}, \citenamefont
  {Gimeno-Segovia}, \citenamefont {Johnston}, \citenamefont {Kieling},
  \citenamefont {Nickerson}, \citenamefont {Pant}, \citenamefont {Pastawski},
  \citenamefont {Rudolph},\ and\ \citenamefont {Sparrow}}]{FBQC2023}%
  \BibitemOpen
  \bibfield  {author} {\bibinfo {author} {\bibfnamefont {S.}~\bibnamefont
  {Bartolucci}}, \bibinfo {author} {\bibfnamefont {P.}~\bibnamefont
  {Birchall}}, \bibinfo {author} {\bibfnamefont {H.}~\bibnamefont
  {Bomb{\'\i}n}}, \bibinfo {author} {\bibfnamefont {H.}~\bibnamefont {Cable}},
  \bibinfo {author} {\bibfnamefont {C.}~\bibnamefont {Dawson}}, \bibinfo
  {author} {\bibfnamefont {M.}~\bibnamefont {Gimeno-Segovia}}, \bibinfo
  {author} {\bibfnamefont {E.}~\bibnamefont {Johnston}}, \bibinfo {author}
  {\bibfnamefont {K.}~\bibnamefont {Kieling}}, \bibinfo {author} {\bibfnamefont
  {N.}~\bibnamefont {Nickerson}}, \bibinfo {author} {\bibfnamefont
  {M.}~\bibnamefont {Pant}}, \bibinfo {author} {\bibfnamefont {F.}~\bibnamefont
  {Pastawski}}, \bibinfo {author} {\bibfnamefont {T.}~\bibnamefont {Rudolph}},
  \ and\ \bibinfo {author} {\bibfnamefont {C.}~\bibnamefont {Sparrow}},\ }\href
  {\doibase 10.1038/s41467-023-36493-1} {\bibfield  {journal} {\bibinfo
  {journal} {Nature Communications}\ }\textbf {\bibinfo {volume} {14}},\
  \bibinfo {pages} {912} (\bibinfo {year} {2023})}\BibitemShut {NoStop}%
\bibitem [{\citenamefont {Giewont}\ \emph {et~al.}(2019)\citenamefont
  {Giewont}, \citenamefont {Nummy}, \citenamefont {Anderson}, \citenamefont
  {Ayala}, \citenamefont {Barwicz}, \citenamefont {Bian}, \citenamefont
  {Dezfulian}, \citenamefont {Gill}, \citenamefont {Houghton}, \citenamefont
  {Hu}, \citenamefont {Peng}, \citenamefont {Rakowski}, \citenamefont {Rauch},
  \citenamefont {Rosenberg}, \citenamefont {Sahin}, \citenamefont {Stobert},\
  and\ \citenamefont {Stricker}}]{Giewont2019}%
  \BibitemOpen
  \bibfield  {author} {\bibinfo {author} {\bibfnamefont {K.}~\bibnamefont
  {Giewont}}, \bibinfo {author} {\bibfnamefont {K.}~\bibnamefont {Nummy}},
  \bibinfo {author} {\bibfnamefont {F.~A.}\ \bibnamefont {Anderson}}, \bibinfo
  {author} {\bibfnamefont {J.}~\bibnamefont {Ayala}}, \bibinfo {author}
  {\bibfnamefont {T.}~\bibnamefont {Barwicz}}, \bibinfo {author} {\bibfnamefont
  {Y.}~\bibnamefont {Bian}}, \bibinfo {author} {\bibfnamefont {K.~K.}\
  \bibnamefont {Dezfulian}}, \bibinfo {author} {\bibfnamefont {D.~M.}\
  \bibnamefont {Gill}}, \bibinfo {author} {\bibfnamefont {T.}~\bibnamefont
  {Houghton}}, \bibinfo {author} {\bibfnamefont {S.}~\bibnamefont {Hu}},
  \bibinfo {author} {\bibfnamefont {B.}~\bibnamefont {Peng}}, \bibinfo {author}
  {\bibfnamefont {M.}~\bibnamefont {Rakowski}}, \bibinfo {author}
  {\bibfnamefont {S.}~\bibnamefont {Rauch}}, \bibinfo {author} {\bibfnamefont
  {J.~C.}\ \bibnamefont {Rosenberg}}, \bibinfo {author} {\bibfnamefont
  {A.}~\bibnamefont {Sahin}}, \bibinfo {author} {\bibfnamefont
  {I.}~\bibnamefont {Stobert}}, \ and\ \bibinfo {author} {\bibfnamefont
  {A.}~\bibnamefont {Stricker}},\ }\href {\doibase 10.1109/JSTQE.2019.2908790}
  {\bibfield  {journal} {\bibinfo  {journal} {IEEE Journal of Selected Topics
  in Quantum Electronics}\ }\textbf {\bibinfo {volume} {25}},\ \bibinfo {pages}
  {1} (\bibinfo {year} {2019})}\BibitemShut {NoStop}%
\bibitem [{\citenamefont {Siew}\ \emph {et~al.}(2021)\citenamefont {Siew},
  \citenamefont {Li}, \citenamefont {Gao}, \citenamefont {Zheng}, \citenamefont
  {Zhang}, \citenamefont {Guo}, \citenamefont {Xie}, \citenamefont {Song},
  \citenamefont {Dong}, \citenamefont {Luo}, \citenamefont {Li}, \citenamefont
  {Luo},\ and\ \citenamefont {Lo}}]{Siew2020}%
  \BibitemOpen
  \bibfield  {author} {\bibinfo {author} {\bibfnamefont {S.~Y.}\ \bibnamefont
  {Siew}}, \bibinfo {author} {\bibfnamefont {B.}~\bibnamefont {Li}}, \bibinfo
  {author} {\bibfnamefont {F.}~\bibnamefont {Gao}}, \bibinfo {author}
  {\bibfnamefont {H.~Y.}\ \bibnamefont {Zheng}}, \bibinfo {author}
  {\bibfnamefont {W.}~\bibnamefont {Zhang}}, \bibinfo {author} {\bibfnamefont
  {P.}~\bibnamefont {Guo}}, \bibinfo {author} {\bibfnamefont {S.~W.}\
  \bibnamefont {Xie}}, \bibinfo {author} {\bibfnamefont {A.}~\bibnamefont
  {Song}}, \bibinfo {author} {\bibfnamefont {B.}~\bibnamefont {Dong}}, \bibinfo
  {author} {\bibfnamefont {L.~W.}\ \bibnamefont {Luo}}, \bibinfo {author}
  {\bibfnamefont {C.}~\bibnamefont {Li}}, \bibinfo {author} {\bibfnamefont
  {X.}~\bibnamefont {Luo}}, \ and\ \bibinfo {author} {\bibfnamefont {G.-Q.}\
  \bibnamefont {Lo}},\ }\href {\doibase 10.1109/JLT.2021.3066203} {\bibfield
  {journal} {\bibinfo  {journal} {Journal of Lightwave Technology}\ }\textbf
  {\bibinfo {volume} {39}},\ \bibinfo {pages} {4374} (\bibinfo {year}
  {2021})}\BibitemShut {NoStop}%
\bibitem [{\citenamefont {Shekhar}\ \emph {et~al.}(2024)\citenamefont
  {Shekhar}, \citenamefont {Bogaerts}, \citenamefont {Chrostowski},
  \citenamefont {Bowers}, \citenamefont {Hochberg}, \citenamefont {Soref},\
  and\ \citenamefont {Shastri}}]{Shekhar2024}%
  \BibitemOpen
  \bibfield  {author} {\bibinfo {author} {\bibfnamefont {S.}~\bibnamefont
  {Shekhar}}, \bibinfo {author} {\bibfnamefont {W.}~\bibnamefont {Bogaerts}},
  \bibinfo {author} {\bibfnamefont {L.}~\bibnamefont {Chrostowski}}, \bibinfo
  {author} {\bibfnamefont {J.~E.}\ \bibnamefont {Bowers}}, \bibinfo {author}
  {\bibfnamefont {M.}~\bibnamefont {Hochberg}}, \bibinfo {author}
  {\bibfnamefont {R.}~\bibnamefont {Soref}}, \ and\ \bibinfo {author}
  {\bibfnamefont {B.~J.}\ \bibnamefont {Shastri}},\ }\href {\doibase
  10.1038/s41467-024-44750-0} {\bibfield  {journal} {\bibinfo  {journal}
  {Nature Communications}\ }\textbf {\bibinfo {volume} {15}},\ \bibinfo {pages}
  {751} (\bibinfo {year} {2024})}\BibitemShut {NoStop}%
\bibitem [{\citenamefont {Bartolucci}\ \emph
  {et~al.}(2021{\natexlab{a}})\citenamefont {Bartolucci}, \citenamefont
  {Birchall}, \citenamefont {Gimeno-Segovia}, \citenamefont {Johnston},
  \citenamefont {Kieling}, \citenamefont {Pant}, \citenamefont {Rudolph},
  \citenamefont {Smith}, \citenamefont {Sparrow},\ and\ \citenamefont
  {Vidrighin}}]{bartolucci2021}%
  \BibitemOpen
  \bibfield  {author} {\bibinfo {author} {\bibfnamefont {S.}~\bibnamefont
  {Bartolucci}}, \bibinfo {author} {\bibfnamefont {P.~M.}\ \bibnamefont
  {Birchall}}, \bibinfo {author} {\bibfnamefont {M.}~\bibnamefont
  {Gimeno-Segovia}}, \bibinfo {author} {\bibfnamefont {E.}~\bibnamefont
  {Johnston}}, \bibinfo {author} {\bibfnamefont {K.}~\bibnamefont {Kieling}},
  \bibinfo {author} {\bibfnamefont {M.}~\bibnamefont {Pant}}, \bibinfo {author}
  {\bibfnamefont {T.}~\bibnamefont {Rudolph}}, \bibinfo {author} {\bibfnamefont
  {J.}~\bibnamefont {Smith}}, \bibinfo {author} {\bibfnamefont
  {C.}~\bibnamefont {Sparrow}}, \ and\ \bibinfo {author} {\bibfnamefont
  {M.~D.}\ \bibnamefont {Vidrighin}},\ }\href@noop {} {} (\bibinfo {year}
  {2021}{\natexlab{a}}),\ \Eprint {http://arxiv.org/abs/2106.13825}
  {arXiv:2106.13825 [quant-ph]} \BibitemShut {NoStop}%
\bibitem [{\citenamefont {Silverstone}\ \emph {et~al.}(2016)\citenamefont
  {Silverstone}, \citenamefont {Bonneau}, \citenamefont {O’Brien},\ and\
  \citenamefont {Thompson}}]{Silverstone2016}%
  \BibitemOpen
  \bibfield  {author} {\bibinfo {author} {\bibfnamefont {J.~W.}\ \bibnamefont
  {Silverstone}}, \bibinfo {author} {\bibfnamefont {D.}~\bibnamefont
  {Bonneau}}, \bibinfo {author} {\bibfnamefont {J.~L.}\ \bibnamefont
  {O’Brien}}, \ and\ \bibinfo {author} {\bibfnamefont {M.~G.}\ \bibnamefont
  {Thompson}},\ }\href {\doibase 10.1109/JSTQE.2016.2573218} {\bibfield
  {journal} {\bibinfo  {journal} {IEEE Journal of Selected Topics in Quantum
  Electronics}\ }\textbf {\bibinfo {volume} {22}},\ \bibinfo {pages} {390}
  (\bibinfo {year} {2016})}\BibitemShut {NoStop}%
\bibitem [{\citenamefont {Vernon}\ \emph {et~al.}(2017)\citenamefont {Vernon},
  \citenamefont {Menotti}, \citenamefont {Tison}, \citenamefont {Steidle},
  \citenamefont {Fanto}, \citenamefont {Thomas}, \citenamefont {Preble},
  \citenamefont {Smith}, \citenamefont {Alsing}, \citenamefont {Liscidini},\
  and\ \citenamefont {Sipe}}]{Vernon17}%
  \BibitemOpen
  \bibfield  {author} {\bibinfo {author} {\bibfnamefont {Z.}~\bibnamefont
  {Vernon}}, \bibinfo {author} {\bibfnamefont {M.}~\bibnamefont {Menotti}},
  \bibinfo {author} {\bibfnamefont {C.~C.}\ \bibnamefont {Tison}}, \bibinfo
  {author} {\bibfnamefont {J.~A.}\ \bibnamefont {Steidle}}, \bibinfo {author}
  {\bibfnamefont {M.~L.}\ \bibnamefont {Fanto}}, \bibinfo {author}
  {\bibfnamefont {P.~M.}\ \bibnamefont {Thomas}}, \bibinfo {author}
  {\bibfnamefont {S.~F.}\ \bibnamefont {Preble}}, \bibinfo {author}
  {\bibfnamefont {A.~M.}\ \bibnamefont {Smith}}, \bibinfo {author}
  {\bibfnamefont {P.~M.}\ \bibnamefont {Alsing}}, \bibinfo {author}
  {\bibfnamefont {M.}~\bibnamefont {Liscidini}}, \ and\ \bibinfo {author}
  {\bibfnamefont {J.~E.}\ \bibnamefont {Sipe}},\ }\href {\doibase
  10.1364/OL.42.003638} {\bibfield  {journal} {\bibinfo  {journal} {Opt.
  Lett.}\ }\textbf {\bibinfo {volume} {42}},\ \bibinfo {pages} {3638} (\bibinfo
  {year} {2017})}\BibitemShut {NoStop}%
\bibitem [{Sup()}]{SupMat}%
  \BibitemOpen
  \href@noop {} {}\bibinfo {note} {{Supplementary Material (to be published at
  a later date).}}\BibitemShut {Stop}%
\bibitem [{\citenamefont {Hu}\ \emph {et~al.}(2009)\citenamefont {Hu},
  \citenamefont {Holzwarth}, \citenamefont {Masciarelli}, \citenamefont
  {Dauler},\ and\ \citenamefont {Berggren}}]{Hu2009}%
  \BibitemOpen
  \bibfield  {author} {\bibinfo {author} {\bibfnamefont {X.}~\bibnamefont
  {Hu}}, \bibinfo {author} {\bibfnamefont {C.~W.}\ \bibnamefont {Holzwarth}},
  \bibinfo {author} {\bibfnamefont {D.}~\bibnamefont {Masciarelli}}, \bibinfo
  {author} {\bibfnamefont {E.~A.}\ \bibnamefont {Dauler}}, \ and\ \bibinfo
  {author} {\bibfnamefont {K.~K.}\ \bibnamefont {Berggren}},\ }\href {\doibase
  10.1109/TASC.2009.2018035} {\bibfield  {journal} {\bibinfo  {journal} {IEEE
  Transactions on Applied Superconductivity}\ }\textbf {\bibinfo {volume}
  {19}},\ \bibinfo {pages} {336} (\bibinfo {year} {2009})}\BibitemShut
  {NoStop}%
\bibitem [{\citenamefont {Chow}\ \emph {et~al.}(2012)\citenamefont {Chow},
  \citenamefont {Gambetta}, \citenamefont {C\'orcoles}, \citenamefont {Merkel},
  \citenamefont {Smolin}, \citenamefont {Rigetti}, \citenamefont {Poletto},
  \citenamefont {Keefe}, \citenamefont {Rothwell}, \citenamefont {Rozen},
  \citenamefont {Ketchen},\ and\ \citenamefont {Steffen}}]{Chow2015}%
  \BibitemOpen
  \bibfield  {author} {\bibinfo {author} {\bibfnamefont {J.~M.}\ \bibnamefont
  {Chow}}, \bibinfo {author} {\bibfnamefont {J.~M.}\ \bibnamefont {Gambetta}},
  \bibinfo {author} {\bibfnamefont {A.~D.}\ \bibnamefont {C\'orcoles}},
  \bibinfo {author} {\bibfnamefont {S.~T.}\ \bibnamefont {Merkel}}, \bibinfo
  {author} {\bibfnamefont {J.~A.}\ \bibnamefont {Smolin}}, \bibinfo {author}
  {\bibfnamefont {C.}~\bibnamefont {Rigetti}}, \bibinfo {author} {\bibfnamefont
  {S.}~\bibnamefont {Poletto}}, \bibinfo {author} {\bibfnamefont {G.~A.}\
  \bibnamefont {Keefe}}, \bibinfo {author} {\bibfnamefont {M.~B.}\ \bibnamefont
  {Rothwell}}, \bibinfo {author} {\bibfnamefont {J.~R.}\ \bibnamefont {Rozen}},
  \bibinfo {author} {\bibfnamefont {M.~B.}\ \bibnamefont {Ketchen}}, \ and\
  \bibinfo {author} {\bibfnamefont {M.}~\bibnamefont {Steffen}},\ }\href
  {\doibase 10.1103/PhysRevLett.109.060501} {\bibfield  {journal} {\bibinfo
  {journal} {Phys. Rev. Lett.}\ }\textbf {\bibinfo {volume} {109}},\ \bibinfo
  {pages} {060501} (\bibinfo {year} {2012})}\BibitemShut {NoStop}%
\bibitem [{\citenamefont {Bienfang}\ \emph {et~al.}(2023)\citenamefont
  {Bienfang}, \citenamefont {Bienfang}, \citenamefont {Gerrits}, \citenamefont
  {Kuo}, \citenamefont {Migdall}, \citenamefont {Polyakov},\ and\ \citenamefont
  {Slattery}}]{bienfang2023single}%
  \BibitemOpen
  \bibfield  {author} {\bibinfo {author} {\bibfnamefont {J.~C.}\ \bibnamefont
  {Bienfang}}, \bibinfo {author} {\bibfnamefont {J.}~\bibnamefont {Bienfang}},
  \bibinfo {author} {\bibfnamefont {T.}~\bibnamefont {Gerrits}}, \bibinfo
  {author} {\bibfnamefont {P.}~\bibnamefont {Kuo}}, \bibinfo {author}
  {\bibfnamefont {A.}~\bibnamefont {Migdall}}, \bibinfo {author} {\bibfnamefont
  {S.}~\bibnamefont {Polyakov}}, \ and\ \bibinfo {author} {\bibfnamefont
  {O.~T.}\ \bibnamefont {Slattery}},\ }\href
  {https://doi.org/10.6028/NIST.IR.8486} {\bibfield  {journal} {\bibinfo
  {journal} {IR 8486, US Department of Commerce, National Institute of
  Standards and Technology}\ } (\bibinfo {year} {2023})}\BibitemShut {NoStop}%
\bibitem [{\citenamefont {Politi}\ \emph {et~al.}(2008)\citenamefont {Politi},
  \citenamefont {Cryan}, \citenamefont {Rarity}, \citenamefont {Yu},\ and\
  \citenamefont {O'Brien}}]{Politi2008}%
  \BibitemOpen
  \bibfield  {author} {\bibinfo {author} {\bibfnamefont {A.}~\bibnamefont
  {Politi}}, \bibinfo {author} {\bibfnamefont {M.~J.}\ \bibnamefont {Cryan}},
  \bibinfo {author} {\bibfnamefont {J.~G.}\ \bibnamefont {Rarity}}, \bibinfo
  {author} {\bibfnamefont {S.}~\bibnamefont {Yu}}, \ and\ \bibinfo {author}
  {\bibfnamefont {J.~L.}\ \bibnamefont {O'Brien}},\ }\href {\doibase
  10.1126/science.1155441} {\bibfield  {journal} {\bibinfo  {journal}
  {Science}\ }\textbf {\bibinfo {volume} {320}},\ \bibinfo {pages} {646}
  (\bibinfo {year} {2008})}\BibitemShut {NoStop}%
\bibitem [{\citenamefont {Lauk}\ \emph {et~al.}(2020)\citenamefont {Lauk},
  \citenamefont {Sinclair}, \citenamefont {Barzanjeh}, \citenamefont {Covey},
  \citenamefont {Saffman}, \citenamefont {Spiropulu},\ and\ \citenamefont
  {Simon}}]{Lauk2020}%
  \BibitemOpen
  \bibfield  {author} {\bibinfo {author} {\bibfnamefont {N.}~\bibnamefont
  {Lauk}}, \bibinfo {author} {\bibfnamefont {N.}~\bibnamefont {Sinclair}},
  \bibinfo {author} {\bibfnamefont {S.}~\bibnamefont {Barzanjeh}}, \bibinfo
  {author} {\bibfnamefont {J.~P.}\ \bibnamefont {Covey}}, \bibinfo {author}
  {\bibfnamefont {M.}~\bibnamefont {Saffman}}, \bibinfo {author} {\bibfnamefont
  {M.}~\bibnamefont {Spiropulu}}, \ and\ \bibinfo {author} {\bibfnamefont
  {C.}~\bibnamefont {Simon}},\ }\href {\doibase 10.1088/2058-9565/ab788a}
  {\bibfield  {journal} {\bibinfo  {journal} {Quantum Science and Technology}\
  }\textbf {\bibinfo {volume} {5}},\ \bibinfo {pages} {020501} (\bibinfo {year}
  {2020})}\BibitemShut {NoStop}%
\bibitem [{\citenamefont {Bombin}\ \emph {et~al.}(2021)\citenamefont {Bombin},
  \citenamefont {Kim}, \citenamefont {Litinski}, \citenamefont {Nickerson},
  \citenamefont {Pant}, \citenamefont {Pastawski}, \citenamefont {Roberts},\
  and\ \citenamefont {Rudolph}}]{Interleaving2021}%
  \BibitemOpen
  \bibfield  {author} {\bibinfo {author} {\bibfnamefont {H.}~\bibnamefont
  {Bombin}}, \bibinfo {author} {\bibfnamefont {I.~H.}\ \bibnamefont {Kim}},
  \bibinfo {author} {\bibfnamefont {D.}~\bibnamefont {Litinski}}, \bibinfo
  {author} {\bibfnamefont {N.}~\bibnamefont {Nickerson}}, \bibinfo {author}
  {\bibfnamefont {M.}~\bibnamefont {Pant}}, \bibinfo {author} {\bibfnamefont
  {F.}~\bibnamefont {Pastawski}}, \bibinfo {author} {\bibfnamefont
  {S.}~\bibnamefont {Roberts}}, \ and\ \bibinfo {author} {\bibfnamefont
  {T.}~\bibnamefont {Rudolph}},\ }\href@noop {} {} (\bibinfo {year} {2021}),\
  \Eprint {http://arxiv.org/abs/2103.08612} {arXiv:2103.08612 [quant-ph]}
  \BibitemShut {NoStop}%
\bibitem [{\citenamefont {Litinski}\ and\ \citenamefont
  {Nickerson}(2022)}]{litinski2022active}%
  \BibitemOpen
  \bibfield  {author} {\bibinfo {author} {\bibfnamefont {D.}~\bibnamefont
  {Litinski}}\ and\ \bibinfo {author} {\bibfnamefont {N.}~\bibnamefont
  {Nickerson}},\ }\href@noop {} {} (\bibinfo {year} {2022}),\ \Eprint
  {http://arxiv.org/abs/2211.15465} {arXiv:2211.15465 [quant-ph]} \BibitemShut
  {NoStop}%
\bibitem [{\citenamefont {Llewellyn}\ \emph {et~al.}(2020)\citenamefont
  {Llewellyn}, \citenamefont {Ding}, \citenamefont {Faruque}, \citenamefont
  {Paesani}, \citenamefont {Bacco}, \citenamefont {Santagati}, \citenamefont
  {Qian}, \citenamefont {Li}, \citenamefont {Xiao}, \citenamefont {Huber},
  \citenamefont {Malik}, \citenamefont {Sinclair}, \citenamefont {Zhou},
  \citenamefont {Rottwitt}, \citenamefont {O'Brien}, \citenamefont {Rarity},
  \citenamefont {Gong}, \citenamefont {Oxenlowe}, \citenamefont {Wang},\ and\
  \citenamefont {Thompson}}]{Llewellyn2020}%
  \BibitemOpen
  \bibfield  {author} {\bibinfo {author} {\bibfnamefont {D.}~\bibnamefont
  {Llewellyn}}, \bibinfo {author} {\bibfnamefont {Y.}~\bibnamefont {Ding}},
  \bibinfo {author} {\bibfnamefont {I.~I.}\ \bibnamefont {Faruque}}, \bibinfo
  {author} {\bibfnamefont {S.}~\bibnamefont {Paesani}}, \bibinfo {author}
  {\bibfnamefont {D.}~\bibnamefont {Bacco}}, \bibinfo {author} {\bibfnamefont
  {R.}~\bibnamefont {Santagati}}, \bibinfo {author} {\bibfnamefont {Y.-J.}\
  \bibnamefont {Qian}}, \bibinfo {author} {\bibfnamefont {Y.}~\bibnamefont
  {Li}}, \bibinfo {author} {\bibfnamefont {Y.-F.}\ \bibnamefont {Xiao}},
  \bibinfo {author} {\bibfnamefont {M.}~\bibnamefont {Huber}}, \bibinfo
  {author} {\bibfnamefont {M.}~\bibnamefont {Malik}}, \bibinfo {author}
  {\bibfnamefont {G.~F.}\ \bibnamefont {Sinclair}}, \bibinfo {author}
  {\bibfnamefont {X.}~\bibnamefont {Zhou}}, \bibinfo {author} {\bibfnamefont
  {K.}~\bibnamefont {Rottwitt}}, \bibinfo {author} {\bibfnamefont {J.~L.}\
  \bibnamefont {O'Brien}}, \bibinfo {author} {\bibfnamefont {J.~G.}\
  \bibnamefont {Rarity}}, \bibinfo {author} {\bibfnamefont {Q.}~\bibnamefont
  {Gong}}, \bibinfo {author} {\bibfnamefont {L.~K.}\ \bibnamefont {Oxenlowe}},
  \bibinfo {author} {\bibfnamefont {J.}~\bibnamefont {Wang}}, \ and\ \bibinfo
  {author} {\bibfnamefont {M.~G.}\ \bibnamefont {Thompson}},\ }\href {\doibase
  10.1038/s41567-019-0727-x} {\bibfield  {journal} {\bibinfo  {journal} {Nature
  Physics}\ }\textbf {\bibinfo {volume} {16}},\ \bibinfo {pages} {148}
  (\bibinfo {year} {2020})}\BibitemShut {NoStop}%
\bibitem [{\citenamefont {Jahanmirinejad}\ and\ \citenamefont
  {Fiore}(2012)}]{Jahanmirinejad:12}%
  \BibitemOpen
  \bibfield  {author} {\bibinfo {author} {\bibfnamefont {S.}~\bibnamefont
  {Jahanmirinejad}}\ and\ \bibinfo {author} {\bibfnamefont {A.}~\bibnamefont
  {Fiore}},\ }\href {\doibase 10.1364/OE.20.005017} {\bibfield  {journal}
  {\bibinfo  {journal} {Opt. Express}\ }\textbf {\bibinfo {volume} {20}},\
  \bibinfo {pages} {5017} (\bibinfo {year} {2012})}\BibitemShut {NoStop}%
\bibitem [{\citenamefont {Sacher}\ \emph {et~al.}(2018)\citenamefont {Sacher},
  \citenamefont {Mikkelsen}, \citenamefont {Huang}, \citenamefont {Mak},
  \citenamefont {Yong}, \citenamefont {Luo}, \citenamefont {Li}, \citenamefont
  {Dumais}, \citenamefont {Jiang}, \citenamefont {Goodwill} \emph
  {et~al.}}]{sacher2018}%
  \BibitemOpen
  \bibfield  {author} {\bibinfo {author} {\bibfnamefont {W.~D.}\ \bibnamefont
  {Sacher}}, \bibinfo {author} {\bibfnamefont {J.~C.}\ \bibnamefont
  {Mikkelsen}}, \bibinfo {author} {\bibfnamefont {Y.}~\bibnamefont {Huang}},
  \bibinfo {author} {\bibfnamefont {J.~C.}\ \bibnamefont {Mak}}, \bibinfo
  {author} {\bibfnamefont {Z.}~\bibnamefont {Yong}}, \bibinfo {author}
  {\bibfnamefont {X.}~\bibnamefont {Luo}}, \bibinfo {author} {\bibfnamefont
  {Y.}~\bibnamefont {Li}}, \bibinfo {author} {\bibfnamefont {P.}~\bibnamefont
  {Dumais}}, \bibinfo {author} {\bibfnamefont {J.}~\bibnamefont {Jiang}},
  \bibinfo {author} {\bibfnamefont {D.}~\bibnamefont {Goodwill}},  \emph
  {et~al.},\ }\href@noop {} {\bibfield  {journal} {\bibinfo  {journal}
  {Proceedings of the IEEE}\ }\textbf {\bibinfo {volume} {106}},\ \bibinfo
  {pages} {2232} (\bibinfo {year} {2018})}\BibitemShut {NoStop}%
\bibitem [{\citenamefont {Bristow}\ \emph {et~al.}(2007)\citenamefont
  {Bristow}, \citenamefont {Rotenberg},\ and\ \citenamefont {van
  Driel}}]{Bristow2007}%
  \BibitemOpen
  \bibfield  {author} {\bibinfo {author} {\bibfnamefont {A.~D.}\ \bibnamefont
  {Bristow}}, \bibinfo {author} {\bibfnamefont {N.}~\bibnamefont {Rotenberg}},
  \ and\ \bibinfo {author} {\bibfnamefont {H.~M.}\ \bibnamefont {van Driel}},\
  }\href {\doibase 10.1063/1.2737359} {\bibfield  {journal} {\bibinfo
  {journal} {Applied Physics Letters}\ }\textbf {\bibinfo {volume} {90}},\
  \bibinfo {pages} {191104} (\bibinfo {year} {2007})}\BibitemShut {NoStop}%
\bibitem [{\citenamefont {Rosenfeld}\ \emph {et~al.}(2020)\citenamefont
  {Rosenfeld}, \citenamefont {Sulway}, \citenamefont {Sinclair}, \citenamefont
  {Anant}, \citenamefont {Thompson}, \citenamefont {Rarity},\ and\
  \citenamefont {Silverstone}}]{Rosenfeld2020}%
  \BibitemOpen
  \bibfield  {author} {\bibinfo {author} {\bibfnamefont {L.~M.}\ \bibnamefont
  {Rosenfeld}}, \bibinfo {author} {\bibfnamefont {D.~A.}\ \bibnamefont
  {Sulway}}, \bibinfo {author} {\bibfnamefont {G.~F.}\ \bibnamefont
  {Sinclair}}, \bibinfo {author} {\bibfnamefont {V.}~\bibnamefont {Anant}},
  \bibinfo {author} {\bibfnamefont {M.~G.}\ \bibnamefont {Thompson}}, \bibinfo
  {author} {\bibfnamefont {J.~G.}\ \bibnamefont {Rarity}}, \ and\ \bibinfo
  {author} {\bibfnamefont {J.~W.}\ \bibnamefont {Silverstone}},\ }\href
  {\doibase 10.1364/OE.386615} {\bibfield  {journal} {\bibinfo  {journal} {Opt.
  Express}\ }\textbf {\bibinfo {volume} {28}},\ \bibinfo {pages} {37092}
  (\bibinfo {year} {2020})}\BibitemShut {NoStop}%
\bibitem [{\citenamefont {Bartolucci}\ \emph
  {et~al.}(2021{\natexlab{b}})\citenamefont {Bartolucci}, \citenamefont
  {Birchall}, \citenamefont {Bonneau}, \citenamefont {Cable}, \citenamefont
  {Gimeno-Segovia}, \citenamefont {Kieling}, \citenamefont {Nickerson},
  \citenamefont {Rudolph},\ and\ \citenamefont {Sparrow}}]{SwitchNetworks2021}%
  \BibitemOpen
  \bibfield  {author} {\bibinfo {author} {\bibfnamefont {S.}~\bibnamefont
  {Bartolucci}}, \bibinfo {author} {\bibfnamefont {P.}~\bibnamefont
  {Birchall}}, \bibinfo {author} {\bibfnamefont {D.}~\bibnamefont {Bonneau}},
  \bibinfo {author} {\bibfnamefont {H.}~\bibnamefont {Cable}}, \bibinfo
  {author} {\bibfnamefont {M.}~\bibnamefont {Gimeno-Segovia}}, \bibinfo
  {author} {\bibfnamefont {K.}~\bibnamefont {Kieling}}, \bibinfo {author}
  {\bibfnamefont {N.}~\bibnamefont {Nickerson}}, \bibinfo {author}
  {\bibfnamefont {T.}~\bibnamefont {Rudolph}}, \ and\ \bibinfo {author}
  {\bibfnamefont {C.}~\bibnamefont {Sparrow}},\ }\href@noop {} {} (\bibinfo
  {year} {2021}{\natexlab{b}}),\ \Eprint {http://arxiv.org/abs/2109.13760}
  {arXiv:2109.13760 [quant-ph]} \BibitemShut {NoStop}%
\bibitem [{\citenamefont {Eltes}\ \emph {et~al.}(2020)\citenamefont {Eltes},
  \citenamefont {Villarreal-Garcia}, \citenamefont {Caimi}, \citenamefont
  {Siegwart}, \citenamefont {Gentile}, \citenamefont {Hart}, \citenamefont
  {Stark}, \citenamefont {Marshall}, \citenamefont {Thompson}, \citenamefont
  {Barreto}, \citenamefont {Fompeyrine},\ and\ \citenamefont
  {Abel}}]{Eltes2020}%
  \BibitemOpen
  \bibfield  {author} {\bibinfo {author} {\bibfnamefont {F.}~\bibnamefont
  {Eltes}}, \bibinfo {author} {\bibfnamefont {G.~E.}\ \bibnamefont
  {Villarreal-Garcia}}, \bibinfo {author} {\bibfnamefont {D.}~\bibnamefont
  {Caimi}}, \bibinfo {author} {\bibfnamefont {H.}~\bibnamefont {Siegwart}},
  \bibinfo {author} {\bibfnamefont {A.~A.}\ \bibnamefont {Gentile}}, \bibinfo
  {author} {\bibfnamefont {A.}~\bibnamefont {Hart}}, \bibinfo {author}
  {\bibfnamefont {P.}~\bibnamefont {Stark}}, \bibinfo {author} {\bibfnamefont
  {G.~D.}\ \bibnamefont {Marshall}}, \bibinfo {author} {\bibfnamefont {M.~G.}\
  \bibnamefont {Thompson}}, \bibinfo {author} {\bibfnamefont {J.}~\bibnamefont
  {Barreto}}, \bibinfo {author} {\bibfnamefont {J.}~\bibnamefont {Fompeyrine}},
  \ and\ \bibinfo {author} {\bibfnamefont {S.}~\bibnamefont {Abel}},\ }\href
  {\doibase 10.1038/s41563-020-0725-5} {\bibfield  {journal} {\bibinfo
  {journal} {Nature Materials}\ }\textbf {\bibinfo {volume} {19}},\ \bibinfo
  {pages} {1164} (\bibinfo {year} {2020})}\BibitemShut {NoStop}%
\bibitem [{\citenamefont {Zhu}\ \emph {et~al.}(2021)\citenamefont {Zhu},
  \citenamefont {Shao}, \citenamefont {Yu}, \citenamefont {Cheng},
  \citenamefont {Desiatov}, \citenamefont {Xin}, \citenamefont {Hu},
  \citenamefont {Holzgrafe}, \citenamefont {Ghosh}, \citenamefont
  {Shams-Ansari}, \citenamefont {Puma}, \citenamefont {Sinclair}, \citenamefont
  {Reimer}, \citenamefont {Zhang},\ and\ \citenamefont {Lon\v{c}ar}}]{Zhu21}%
  \BibitemOpen
  \bibfield  {author} {\bibinfo {author} {\bibfnamefont {D.}~\bibnamefont
  {Zhu}}, \bibinfo {author} {\bibfnamefont {L.}~\bibnamefont {Shao}}, \bibinfo
  {author} {\bibfnamefont {M.}~\bibnamefont {Yu}}, \bibinfo {author}
  {\bibfnamefont {R.}~\bibnamefont {Cheng}}, \bibinfo {author} {\bibfnamefont
  {B.}~\bibnamefont {Desiatov}}, \bibinfo {author} {\bibfnamefont {C.~J.}\
  \bibnamefont {Xin}}, \bibinfo {author} {\bibfnamefont {Y.}~\bibnamefont
  {Hu}}, \bibinfo {author} {\bibfnamefont {J.}~\bibnamefont {Holzgrafe}},
  \bibinfo {author} {\bibfnamefont {S.}~\bibnamefont {Ghosh}}, \bibinfo
  {author} {\bibfnamefont {A.}~\bibnamefont {Shams-Ansari}}, \bibinfo {author}
  {\bibfnamefont {E.}~\bibnamefont {Puma}}, \bibinfo {author} {\bibfnamefont
  {N.}~\bibnamefont {Sinclair}}, \bibinfo {author} {\bibfnamefont
  {C.}~\bibnamefont {Reimer}}, \bibinfo {author} {\bibfnamefont
  {M.}~\bibnamefont {Zhang}}, \ and\ \bibinfo {author} {\bibfnamefont
  {M.}~\bibnamefont {Lon\v{c}ar}},\ }\href {\doibase 10.1364/AOP.411024}
  {\bibfield  {journal} {\bibinfo  {journal} {Adv. Opt. Photon.}\ }\textbf
  {\bibinfo {volume} {13}},\ \bibinfo {pages} {242} (\bibinfo {year}
  {2021})}\BibitemShut {NoStop}%
\bibitem [{\citenamefont {Bombin}\ \emph
  {et~al.}(2023{\natexlab{a}})\citenamefont {Bombin}, \citenamefont {Dawson},
  \citenamefont {Nickerson}, \citenamefont {Pant},\ and\ \citenamefont
  {Sullivan}}]{bombin2023increasing}%
  \BibitemOpen
  \bibfield  {author} {\bibinfo {author} {\bibfnamefont {H.}~\bibnamefont
  {Bombin}}, \bibinfo {author} {\bibfnamefont {C.}~\bibnamefont {Dawson}},
  \bibinfo {author} {\bibfnamefont {N.}~\bibnamefont {Nickerson}}, \bibinfo
  {author} {\bibfnamefont {M.}~\bibnamefont {Pant}}, \ and\ \bibinfo {author}
  {\bibfnamefont {J.}~\bibnamefont {Sullivan}},\ }\href@noop {} {} (\bibinfo
  {year} {2023}{\natexlab{a}}),\ \Eprint {http://arxiv.org/abs/2303.16122}
  {arXiv:2303.16122 [quant-ph]} \BibitemShut {NoStop}%
\bibitem [{\citenamefont {Bombin}\ \emph
  {et~al.}(2023{\natexlab{b}})\citenamefont {Bombin}, \citenamefont {Dawson},
  \citenamefont {Farrelly}, \citenamefont {Liu}, \citenamefont {Nickerson},
  \citenamefont {Pant}, \citenamefont {Pastawski},\ and\ \citenamefont
  {Roberts}}]{bombin2023_2}%
  \BibitemOpen
  \bibfield  {author} {\bibinfo {author} {\bibfnamefont {H.}~\bibnamefont
  {Bombin}}, \bibinfo {author} {\bibfnamefont {C.}~\bibnamefont {Dawson}},
  \bibinfo {author} {\bibfnamefont {T.}~\bibnamefont {Farrelly}}, \bibinfo
  {author} {\bibfnamefont {Y.}~\bibnamefont {Liu}}, \bibinfo {author}
  {\bibfnamefont {N.}~\bibnamefont {Nickerson}}, \bibinfo {author}
  {\bibfnamefont {M.}~\bibnamefont {Pant}}, \bibinfo {author} {\bibfnamefont
  {F.}~\bibnamefont {Pastawski}}, \ and\ \bibinfo {author} {\bibfnamefont
  {S.}~\bibnamefont {Roberts}},\ }\href@noop {} {} (\bibinfo {year}
  {2023}{\natexlab{b}}),\ \Eprint {http://arxiv.org/abs/2308.07844}
  {arXiv:2308.07844 [quant-ph]} \BibitemShut {NoStop}%
\bibitem [{MIP()}]{MIP}%
  \BibitemOpen
  \href@noop {} {}\bibinfo {note} {{PsiQuantum, Manuscript in
  preparation.}}\BibitemShut {Stop}%
\end{thebibliography}
\end{document}